\documentclass[10pt,journal,compsoc]{IEEEtran}
\usepackage[T1]{fontenc}
\usepackage{booktabs} 
\usepackage{multirow}
\usepackage{makecell} 
\usepackage{tabularx}
\usepackage{graphicx}
\usepackage{url}
\usepackage{mdwlist}
\usepackage{subcaption}
\usepackage{hyperref}
\usepackage{amsfonts}
\usepackage{array}
\usepackage[linesnumbered,ruled]{algorithm2e}
\usepackage{xcolor}
\usepackage{amsmath}
\usepackage{amssymb}
\usepackage[flushleft]{threeparttable}
\usepackage{enumitem}
\usepackage{balance}
\usepackage{dsfont}
\usepackage{verbatim} 

\hypersetup{colorlinks=true,linkcolor=green,urlcolor=blue,citecolor=red}

\newcommand{\wule}[1]{\textcolor{red}{~L:[#1]}}

\newcommand{\wx}[1]{#1}
\newcommand{\zk}[1]{#1}

\newcommand\norm[1]{\left\lVert#1\right\rVert}

\newcommand{\Trans}[1]{{#1}^{\top}}

\newcommand{\Mat}[1]{\mathbf{#1}}

\newcommand{\Space}[1]{\mathbb{#1}}
\newcommand{\Set}[1]{\mathcal{#1}}

\newcommand{\ie}{\emph{i.e., }}

\ifCLASSOPTIONcompsoc
  \usepackage[nocompress]{cite}
\else
  \usepackage{cite}
\fi
\newcommand{\tabincell}[2]{\begin{tabular}{@{}#1@{}}#2\end{tabular}} 

\begin{document}


\title{A Survey on Accuracy-oriented Neural Recommendation: From Collaborative Filtering to Information-rich Recommendation}

\author{Le~Wu~\IEEEmembership{Member,~IEEE}, Xiangnan~He~\IEEEmembership{Member,~IEEE}, 
Xiang Wang~\IEEEmembership{Member,~IEEE}, 
Kun~Zhang~\IEEEmembership{Member,~IEEE},
	and Meng Wang,~\IEEEmembership{Fellow,~IEEE}
	\IEEEcompsocitemizethanks{\IEEEcompsocthanksitem L. Wu, and M Wang are with Key Laboratory of Knowledge Engineering with Big Data, Hefei University of Technology, Hefei, Anhui 230029, China, and Institute of Artificial Intelligence, Hefei Comprehensive National
    Science Center, Hefei, Anhui 230088 (email: lewu.ustc,  eric.mengwang@gmail.com).
    
    \IEEEcompsocthanksitem K. Zhang is with Key Laboratory of Knowledge Engineering with Big Data, Hefei University of Technology, Hefei, Anhui 230029, China.(email:zhang1028kun@gmail.com).

	\IEEEcompsocthanksitem X. He is with  University of Science and Technology of China, Hefei 230026, China. (email: xiangnanhe@gmail.com).
	
	\IEEEcompsocthanksitem X. Wang is with National University of Singapore, Singapore. (email: xiangwang@u.nus.edu).}
	
}

\markboth{IEEE Transactions on Knowledge and Data Engineering}%
{Wu \MakeLowercase{\textit{et al.}}: A Survey on Accuracy-based Neural Recommendation: From Collaborative Filtering to Information-rich Recommendation}

\IEEEtitleabstractindextext{%
\begin{abstract}
Influenced by the \wx{great} success of deep learning in computer vision and language understanding, research in recommendation has shifted to inventing new recommender models based on neural networks. In recent years, we have witnessed significant progress in developing neural recommender models, which generalize and surpass traditional recommender models owing to the strong representation power of neural networks. In this survey paper, we conduct a systematic review on neural recommender models \wx{from the perspective of recommendation modeling with the accuracy goal}, aiming to summarize this field to \wx{facilitate researchers and practitioners working on recommender systems.}
Specifically, \zk{based on the data usage during recommendation modeling, we divide the work into collaborative filtering and information-rich recommendation}: 1) \textit{collaborative filtering}, which leverages the key source of user-item interaction data; 2) \textit{content enriched recommendation}, which additionally utilizes the side information associated with users and items, like user profile and item knowledge graph; and 3) \textit{temporal/sequential recommendation}, which accounts for the contextual information associated with an interaction, such as time, location, and the past interactions.
After reviewing representative work for each type, we finally discuss some promising directions in this field.

\end{abstract}

}

\maketitle

\IEEEdisplaynontitleabstractindextext

\IEEEpeerreviewmaketitle

\IEEEraisesectionheading{\section{Introduction}\label{sec:introduction}}
\IEEEPARstart{I}{nformation} overload is an increasing problem in people's every life due to the proliferation of the Internet.
Recommender system serves as an effective solution to alleviate the information overload issue, to facilitate users seeking desired information, and to increase the traffic and revenue of service providers. 
It has been used in a wide range of applications, such as e-commerce, social media sites, news portals, app stores, digital libraries, and so on. 
It is one of the most ubiquitous user-centered artificial intelligence applications in modern information systems.

The research in recommendation can be dated back to 1990s~\cite{Goldberg:1992}, in the age the early work has developed many heuristics for content-based and Collaborative Filtering (CF)~\cite{TKDE2005Survey}. 
Popularized by the Netflix challenge, Matrix Factorization (MF)~\cite{koren2009matrix} later becomes the mainstream recommender model for a long time~(from 2008 until 2016)~\cite{karatzoglou2010multiverse,FPMC}. However, the linear nature of factorization models makes them less effective when dealing with large and complex data, e.g., the complex user-item interactions, and the items may contain complex semantics (e.g., texts and images) that require a thorough understanding. 
Around the same time in the mid-2010s, the rise of deep neural networks in machine learning (a.k.a., Deep Learning) has revolutionized several areas including speech recognition, computer vision, and natural language processing~\cite{goodfellow2016deep}. The great success of deep learning stems from the considerable expressiveness of neural networks, which are particularly advantageous for learning from large data with complicated patterns. This naturally brings new opportunities to advance the recommendation technologies.  And not surprisingly, there emerges a lot of work on developing neural network approaches to recommender systems in the past several years. In this work, we aim to provide a systematic review on the recommender models that use neural networks --- referred to as ``\textit{neural recommender models}''. This is the most thriving topic in current recommendation research, not only has many exciting progresses in recent years, but also shows the potential to be the technical foundations of the next-generation recommender systems.

\wx{
\subsection{Differences with Existing Surveys.}
Given the significance and popularity of recommendation research, there are some recently published surveys also reviewed this area \cite{TKDE2005Survey,ACM2014CFSurvey,ACM2017CrossSurvey,zhang2018explainable,wang2021graph}. Here we shortly discuss the main differences with these work to highlight the necessity and significance of this survey.

}

\zk{
Existing surveys consist of two main parts. 
The first part focuses on the specific topics or directions, such as side information utilization in collaborative filtering~\cite{ACM2014CFSurvey}, cross-domain recommendation~\cite{ACM2017CrossSurvey}, explainable recommendation~\cite{TEM}, knowledge graph-enhanced recommendation~\cite{guo2020survey}, sequential recommendation \cite{DBLP:journals/tois/FangZSG20,DBLP:journals/csur/QuadranaCJ18}, and session-based recommendation \cite{DBLP:journals/corr/abs-1902-04864}.
The other part follows the taxonomy of Deep Learning~(DL) to summarize the recommendation methods. For example, Zhang et al.\cite{DBLP:journals/csur/ZhangYST19} organized the discussions on recommendation methods into MLP based, autoencoder based, RNN based, attention based, etc. Similar surveys can also be found~\cite{DBLP:journals/air/BatmazYBK19,DBLP:journals/air/DauS20}. These surveys mainly compare the technical difference of using various deep learning methods for recommendation.

Different from existing surveys, our survey is organized from the perspective of recommendation modeling with the accuracy goal, and covers the most typical recommendation scenarios, such as CF, content-enriched methods, and temporal/sequential methods. This will not only help researchers understanding why and when a deep learning technique would work but also facilitate practitioners designing better solutions for a specific recommendation scenario.
}

\zk{
\subsection{How Do We Collect the Papers?}
Since our survey focuses on reviewing recommender system from the perspective of recommendation modeling with the accuracy goal, we retrieved most of the related top conferences such as WWW, SIGIR, KDD, ICLR, AAAI, IJCAI, WSDM, and RecSys, as well as the top journals such as TKDE, TKDD, and so on. 
Meanwhile, we also leveraged Google Scholar to search the recent related work. 
According to the categories that we made in this survey, we used key words such as \textit{collaborative filtering}, \textit{content+RS}, \textit{recommender systems}, \textit{context+RS}, \textit{side information}, \textit{graph neural network}, \textit{neural recommendation}, etc, to search the relevant work. 
Then, based on the retrieved papers, we carefully design the topical structure to cover all papers as completely as possible. 
Besides, in order to avoid missing some important work, we also double-checked those classic and influential papers in recommendation.  
}

\begin{figure}[t]
	\includegraphics[width=\linewidth]{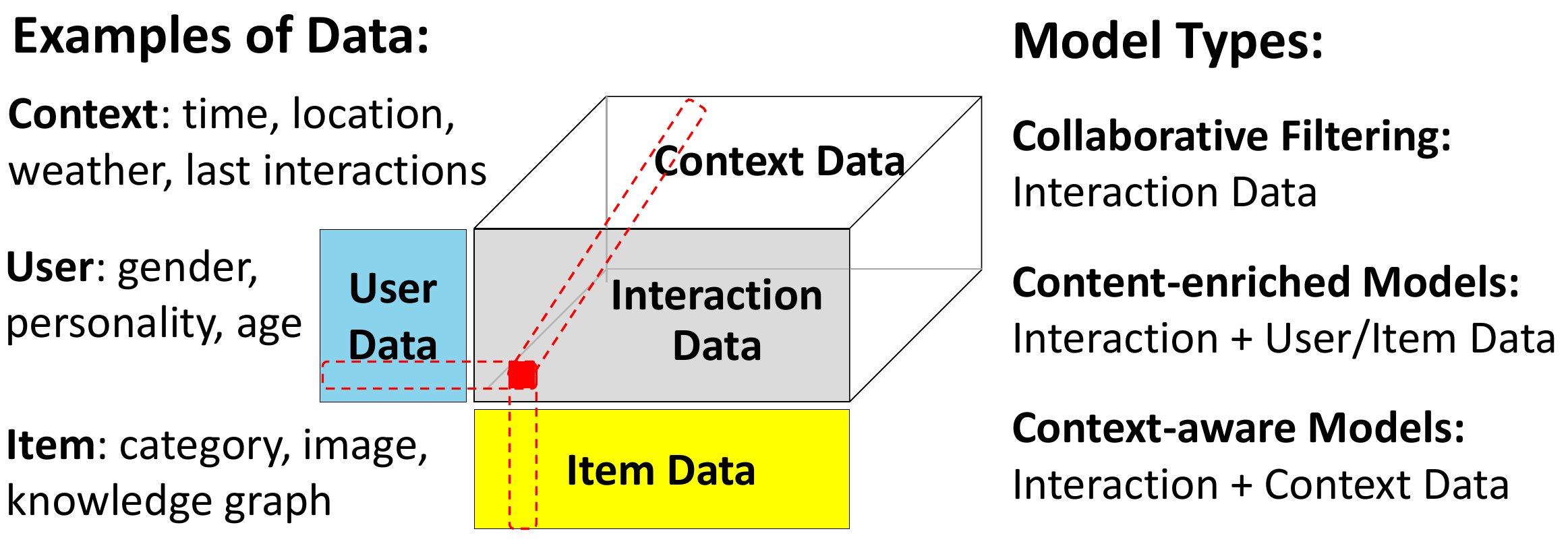}
	\caption{An illustration of the data used for recommendation modeling and the three model types.}
	\vspace{-10pt}
	\label{fig:data}
\end{figure}

\wx{
\subsection{Scope and Organization of This Survey}
This survey is organized into two major parts: Sections \ref{sec:cf} to \ref{sec:contextRS} review existing methods, and Section \ref{sec:conclusion} discusses future directions and open issues.
Before elaborating each section, we first give the problem formulation.

Regardless of the recommendation domain and scenario, we can abstract the ``learning to recommend'' problem as:
\begin{gather}
    \hat{y}_{u,i,c}=f(D_{u}, D_{i}, D_{c}),
\end{gather}
that is, learning the prediction function $f$ to estimate the likelihood that a user $u$ will favor an item $i$ under the context $c$, given the data $D_u$ , $D_i$ , and $D_c$ to describe the user $u$, item $i$, and context $c$, respectively. In doing so, we allow a unified framework to summarize neural recommendation models:
\begin{itemize}
    \item Section \ref{sec:cf} reviews \emph{collaborative filtering models}, which forms the basis of personalized recommendation and is the most researched topic in recommendation. They can be seen as ignoring the context data $D_c$ and using only the ID or interaction history in $D_u$ and $D_i$.
    \item Section \ref{sec:content} reviews the models that integrate the side information of users and items into recommendation, such as user profiles and social network, item attributes and knowledge graph. We term them as \emph{content-enriched models}, which naturally extend collaborative filtering (CF) by integrating the side information into $D_u$ and $D_i$, whereas the context data $D_c$ is also ignored.
    \item Section \ref{sec:contextRS} reviews the models that use contextual information. The contextual data are associated with each user-item interaction, but do not belong to either user content or item content, like time, location, and the past interaction sequence~\cite{TKDE2005Survey}. The \emph{context-aware models} make predictions based on the context data $D_c$, in addition to the user-related data $D_u$ and item-related data $D_i$.
    Due to page limit, we focus on temporal context, which is one of the most common contextual data.
\end{itemize}

Fig.~\ref{fig:data} illustrates the typical data used for recommendation modeling and three model types.
It is worth noting that different models are designed for different recommendation scenarios. Nevertheless, in many cases we can make simple adjustments on a model's component to make it suitable (at least technically viable) for another scenario. For example, many CF models are designed to first obtain user and item representations, and then the prediction function is learned given the user and item representations. To make them be content-enriched, we simply need to enhance the representation learning component with content modeling. Another example is that we can treat the contextual information as part of user data, i.e., constructing $D_{u,c}$ to replace $D_u$, to tweak content-enriched models to also be context-aware.
Although these adjusted models may not be officially proposed or published, they can be obtained without much effort and worth exploring in real applications.
Such design flexibility can be attributed to the layer-wise architecture of neural recommendation models, where different layers are designed for different aims. For convenience,  we also summarize related neural recommendation models into the taxonomy of recommendation modeling \footnote{~\url{https://github.com/lmcRS/AWS-recommendation-papers}}.}
We hope this survey would provide a clear road-map to facilitate practitioners understanding and better designing models for their purpose.


\section{Collaborative Filtering Models}\label{sec:cf}


The concept of CF stems from the idea that leveraging collaborative behaviors of all users for predicting the behavior of a target user. Early approaches directly calculate the behavior similarity of users~(user-based CF) or items~(item-based CF) with memory based models. Later on, matrix factorization based models become prevalent by collectively finding the latent spaces that encode user-item interaction matrix~\cite{koren2009matrix,rendle2009bpr}. 
Given the expressive complex modeling power of neural networks, the current solutions for neural CF can be summarized into two categories: representation modeling of users and items, and user-item interaction modeling given the representations. 

\begin{footnotesize}
\begin{table*}
    \caption{Summarization of representation learning approaches for CF}
    \centering
        \begin{tabular}{|c|c|c|}\hline
           Category & Modeling Summarization & Models \\ \hline
           \multirowcell{3}{Classical\\ Matrix\\ Factorization} & \tabincell{c}{User UID~(Free Embed)\\ Item IID~(Free Embed)} & BPR~\cite{rendle2009bpr}, MF~\cite{koren2009matrix} et al.\\  \cline{2-3}
           & \tabincell{c}{User Interacted items~(Free Embed+Heuristic Agg)\\ Item IID~(Free Embed)}  & FISM~\cite{FISM}, PMLAM~\cite{ma2020probabilistic}, pQCF~\cite{lian2020product}, FAWMF~\cite{chen2020fast}\\ \cline{2-3}
           &\tabincell{c}{User Interacted items+UID~(Free Embed+Heuristic Agg)\\ Item IID~(Free Embed)} & SVD++~\cite{SVD++}  \\  \hline
           
           \multirowcell{2}{History\\ Attention}  &\tabincell{c}{User Interacted items~(Free Embed+Heuristic Agg)\\Item IID~(Free Embed)} & NAIS~\cite{he2018nais}  \\  \cline{2-3}
           &\tabincell{c}{ User Interacted items+UID~(Free Embed+Heuristic Agg)\\ Item IID~(Free Embed) }& ACF~\cite{ACF}  \\ \hline 
           
           \multirowcell{2}{Autoencoder \\ Models} & Item Interacted Items~(Non-linear Encoder) &  AutoRec~\cite{AutoRec}, CDAE~\cite{CDAE}, Mult-VAE~\cite{Liang:2018:VAC} et al. \\  \cline{2-3}
           & \tabincell{c}{User Interacted Items~(Non-linear encoder)\\ Item Interacted users~(Non-linear Encoder) }& REAP~\cite{Zhuang:2017:RLP}, CE-VNCF~\cite{luo2020deep}, SW-DAE~\cite{khawar2020learning}  \\ \hline
           
           \multirowcell{2}{Graph \\ Learning} & \tabincell{c}{User UID+Graph~(GNN)\\ Item IID+Graph~(GNN)} & \tabincell{c}{GC-MC~\cite{GC-MC}, NGCF~\cite{NGCF}, SpectralCF~\cite{SpectralCF},\\ NIA-GCN~\cite{sun2020neighbor}, BGCF~\cite{sun2020framework},  DGCF~\cite{wang2020disentangled} et al.}  \\ \cline{2-3}
           & \tabincell{c}{User UID+Graph~(Simplified GNN)\\ Item IID+Graph~(Simplified GNN)} & LR-GCCF~\cite{LRGCCF}, LightGCN~\cite{lightgcn}, DHCF~\cite{ji2020dual}  et al. \\ \hline
        \end{tabular}
    \label{tab:cf_representation}
\end{table*}
\end{footnotesize}

\subsection{Representation Learning}
Let $\Set{U}$ and $\Set{V}$ denote users and items in CF, with $\mathbf{R}\in\mathbb{R}^{M\times N}$ is a user-item interaction behavior matrix. The general objective is to learn a user embedding matrix $\Mat{P}$ and an item embedding matrix $\Mat{Q}$, with $\Mat{p}_{u}$ and $\Mat{q}_{i}$ denote the representation parameters for user $u$  and item $i$, respectively.


In fact, as each user \wx{has} limited behavior compared to the large item set, a key challenge that lies in CF is the sparsity of the user-item interaction behavior for accurate user and item embedding learning. Different kinds of representation learning models vary in input data, as well as the representation modeling techniques given the input data. We divide this section into three categories: 
\textit{history behavior aggregation enhanced models}, \textit{autoencoder based models}, and \textit{graph learning approaches}. For ease of explanation, we list the typical representation learning models in Table~\ref{tab:cf_representation}.

\subsubsection{History Behavior Attention Aggregation Models}\label{sec:cf-history}
By taking the one-hot User ID~(UID), and one-hot Item ID~(IID) as input, classical latent factor models associate each UID $u$ and IID $i$ with a free embedding vector of $\textbf{p}_u$ and $\textbf{q}_i$~\cite{koren2009matrix,rendle2009bpr}. Instead of modeling users with free embeddings, researchers further proposed borrowing users' historical behavior for better user representation modeling. E.g., Factored Item Similarity Model (FISM) pools the interacted item embeddings as a user representation vector~\cite{FISM}, and  SVD++~\cite{SVD++} adds UID embedding $\textbf{p}_u$ with the interaction history embedding (i.e., the FISM user representation) as the final user representation. These models relied on simple linear matrix factorization, and used heuristics or equal weights for the interaction history aggregation.

However, different historical items should contribute differently to model a user's preference. 
Thus, some researchers integrate neural attention mechanism into history representation learning~\cite{ACF,he2018nais,DeepICF}.
One representative work is Attentive Collaborative Filtering (ACF)~\cite{ACF}, which assigns each interacted item with a user-aware attentive weight to indicate its importance to user representation:
\begin{align}
\hat{r}_{ui} = (\textbf{p}_u +  \sum_{j\in\Set{R}_{u}}\alpha(u,j)\Mat{q}_{j})^T \textbf{q}_i,
\end{align}
where $\textbf{p}_u$ is the ID embedding of user $u$, $\Set{R}_{u}$ denotes the items that $u$ has interacted with. $\alpha(u,j)$ is the attentive weight defined as:
\begin{align}
    \alpha(u,j)=\frac{\exp{(\Set{F}(\Mat{p}_{u},\Mat{q}_{j}))}}{\sum_{j'\in\Set{R}_{u}}\exp{(\Set{F}(\Mat{p}_{u},\Mat{q}_{j'}))}},
\end{align}
where  $\Set{F}(\cdot,\cdot)$ is a function that can be implemented as a MLP or simply inner product.

In practice, the influence of a historical item can be dependent on the target item, e.g., the purchase of a phone case is more related to the previous purchase of phone, while the purchase of a pant could be more related to the previous purchase of a shirt. As such, it may be beneficial to have dynamic user representation when considering the prediction on different target items.
To this end, the Neural Attentive Item Similarity model (NAIS) model~\cite{he2018nais} revises the attention mechanism to be target item-aware:
\begin{equation}
\begin{aligned}
\hat{r}_{ui} &= (\sum_{j\in\Set{R}_{u}}\alpha(i,j)\Mat{q}_{j})^T \textbf{q}_i \\
\alpha(i,j) & =\frac{\exp{(\Set{F}(\Mat{q}_{i},\Mat{q}_{j}))}}{[\sum_{k\in\Set{R}_u}\exp{(\Set{F}(\Mat{q}_{i},\Mat{q}_{k}))}]^{\beta} },
\end{aligned}
\end{equation}
where $\alpha(i,j)$ denotes the contribution of historical item $j$ to user representation when predicting a user's preference on target item $i$. $\beta$ is a hyper-parameter between 0 and 1 (e.g., 0.5), for smoothing the interaction histories of different lengths. Similar attention mechanisms have been adopted for representation learning from interaction history, e.g., the Deep Item-based CF model (DeepICF)~\cite{DeepICF} and Deep Interest Network (DIN)~\cite{DIN}.
As such, interaction history contains more information than single user ID and is a suitable choice for representation learning.

\subsubsection{Autoencoder based Representation Learning}
\zk{By utilizing the idea of reconstructing input for a better representation learning}, autoencoder based models take the incomplete user-item matrix as input, and learn a hidden representation of each instance with an encoder, and further with a decoder part that reconstructs the input based on the hidden representation. 
By treating each user's historical records as input, the autoencoder based models learn each user's latent representation with a complex encoder neural network, and feed the learned user representation into a decoder network to output the predicted preference of each user.  An alternative approach is to take each item's rating records from all users as input, and learn the item's latent representation to reconstruct the predicted preference of each item from all users~\cite{CDAE,AutoRec}. 
\zk{Similar to the development of autoencoder}, the extensions of autoencoder based models can also be classified into two categories. The first category leveraged autoencoder variants, and injected denoising autoencoders~\cite{CDAE}, variational autoencoders~\cite{Liang:2018:VAC} into CF. These models can be seen as using complex deep learning techniques for learning either user or item encoders. The second category exploited the duality of users and items in autoencoders, and designed two parallel encoders to learn the user and item representations, and then also use inner product to model users' preferences to items~\cite{Zhuang:2017:RLP}. It is worth pointing out the autoencoder based CF approaches can also be classified as extensions of the historical behavior attention based models, as these approaches adopt deep neural networks for aggregating historical behavior. 
\zk{Therefore, for the sake of simplicity, we have only briefly introduced  autoencoder based models and have not repeated the specific technical details.}

\subsubsection{Graph based Representation Learning}\label{sec:cf-graph}
The CF effects are reflected in interaction histories of multiple users. As such, using collective interaction histories has the potential to improve the representation quality.
From the perspective of user-item interaction graph, the individual interaction history is equivalent to the first-order connectivity of the user. Thus, a natural extension is to mine the higher-order connectivity from the user-item graph structure. For example, the second-order connectivity of a user consists of similar users who have co-interacted with the same items. 
Fortunately, with the success of Graph Neural Networks~(GNNs) for modeling graph structure data in the community~\cite{kipf2017semi}, many prior studies have been proposed to model the user-item bipartite graph structure for neural graph based representation learning. Given the user-item bipartite graph, let $\mathbf{P}^0$ and $\mathbf{Q}^0$ denote the free user latent matrix and item latent matrix as many classical latent factor based models, i.e., the $0^{th}$-order user and item embedding. These neural graph based models iteratively update the $(l+1)^{th}$-order user~(item) embedding as an aggregation of the $l^{th}$-order item~(user) embedding. For instance,  each user $u$'s updated embedding $\Mat{p}_{u}^{(l+1)}$ is calculated as:

\begin{align}
     &\Mat{a}_{u}^{(l+1)}=Agg(\Mat{q}_{j}^{l}|j\in \Set{R}_u),\\
    &\Mat{p}_{u}^{(l+1)}=\rho(\Mat{W}^l[\Mat{p}_{u}^{l},\Mat{a}_{u}^{(l+1)}]),
\end{align}

\noindent where $\Mat{q}_{j}^{l}$ is item $j$'s representation at $l^{th}$ layer, $\Set{R}_u$ denotes items that connect to user $u$ in the user-item bipartite graph. $\Mat{a}_{u}^{(l+1)}$ is the aggregation of connected items' representations in the $l^{th}$ layer, $\Mat{W}^l$ is an embedding transformation matrix that needs to be learned, and $\rho()$ is an activation function.  After that, each user's~(item's) final embedding can be seen as combining each entity's embedding at each layer.

The above steps can be seen as embedding propagation in the user-item bipartite graph. With a predefined layer $L$, the up to $L^{th}$ order sub graph structure is directly encoded in the user and item embedding representation step. For example, SpectralCF utilized the spectral graph convolutions for CF~\cite{SpectralCF}. GC-MC~\cite{GC-MC} and NGCF~\cite{NGCF} modeled the graph convolutions of user-item 
interactions in the original space, and are more effective and efficient in practice. Very recently, researchers argued that these neural graph based CF models differ from the classical GNNs as CF models do not contain any user or item features. Directly borrowing complex steps such as embedding transformation, and non-linear activations in GNNs may not be a good choice. Simplified neural graph CF models, including LR-GCCF~\cite{LRGCCF}, and LightGCN~\cite{lightgcn} have been proposed, which eliminate unnecessary deep learning operations. These simplified neural graph based models show superior performance in practice without the need of carefully chosen activation functions.

\begin{table*}
    \caption{Interaction modeling techniques}
    \centering
        \begin{tabular}{|c|c|c|}\hline
           Category & Modeling Summarization & Models \\ \hline
           Inner Product  & $\hat{r}_{ui}=\Trans{\Mat{p}}_{u}\Mat{q}_{i}$   & Most models\\ \hline
           \multirowcell{4}{Distance \\ Modeling} &  Euclidean distance $d_{ui}=\norm{\Mat{p}_{u}-\Mat{q}_{i}}^{2}_{2}$ &  CML~\cite{CML} \\ \cline{2-3}
           & Nearby translation $\hat{d}_{ui}=\beta_{j}-d(\Mat{q}_{j}+\Mat{p}_{u}, \Mat{q}_{i})$ & TransRec~\cite{recsys2017translation} \\ \cline{2-3}
           & Memory  enhanced Translation $\hat{d}_{ui}=\norm{\Mat{p}_{u}+\Mat{E}-\Mat{q}_{i}}^{2}_{2}$    & LRML~\cite{LRML} \\ 
           
           \cline{2-3}
           & \wx{Distance in Hpyerbolic Space}    & \wx{HyperML}~\cite{DBLP:conf/wsdm/TranT0CL20} \\
           
           \hline
           \multirowcell{3}{Neural \\ Networks} & $\hat{r}_{ui}=MLP(\Mat{p}_{u}||\Mat{q}_i)$   & NCF~\cite{NCF} et al. \\\cline{2-3}
           & $\hat{r}_{ui}=CNN(\Mat{p}_{u}\otimes\Mat{q}_i)$ &ONCF~\cite{ONCF} et al. \\ \cline{2-3}
            & Autoencoder based reconstruction $\norm{\Mat{r}_{i}-dec(enc(\Mat{r}_i))}^{2}_{2}$ & AutoRec~\cite{AutoRec}, CDAE~\cite{CDAE} et al.\\ \hline
        \end{tabular}
    \label{tab:cf_interaction}
\end{table*}

\subsection{Interaction Modeling}
Let $\mathbf{p}_u$ and $\mathbf{q}_i$ denote the learned embeddings of user $u$ and item $i$ from representation models, this component aims at interaction function modeling that estimates the user's preference towards the target item based on their representations. In the following, we describe how to model users' predicted preference, denoted as $\hat{r}_{ui}$ based on the learned embeddings. For ease of explanation, as shown in Table~\ref{tab:cf_interaction}, we summarize three main categories for interaction modeling:  classical inner product based approaches, distance based modeling and neural network based approaches.

Most previous recommendation models relied on the inner product between user embedding and item embedding to estimate the user-item pair score as: $\hat{r}_{ui}=\Trans{\Mat{p}}_{u}\Mat{q}_{i}=\sum_{f=1}^d p_{uf}q_{if}$. 
Despite its great success and simplicity, prior efforts suggest that simply conducting inner product would have two major limitations. First, the triangle inequality is violated~\cite{CML}. That is, inner product only encourages the representations of users and historical items to be similar, but lacks guarantees for the similarity propagation between user-user and item-item relationships. Second, it models the linear interaction, and may fail to capture the complex relationships between users and items~\cite{NCF}.

\subsubsection{Distance based Metrics}
To solve the first issue, a line of research~\cite{CML,recsys2017translation,LRML} borrows ideas from translation principles and uses distance metric as the interaction function.
The inherent triangle inequality assumption plays an important role in helping capture underlying relationships among users and items. For instance, if user $u$ tends to purchase items $i$ and $j$, the representations of $i$ and $j$ should be close in the latent space.

Towards this end, CML~\cite{CML} minimizes the distance $d_{ui}$ between each user-item interaction $<u,i>$ in Euclidean space  as: $d_{ui}=\norm{\Mat{p}_{u}-\Mat{q}_{i}}^{2}_{2}$. Instead of minimizing the distance between each observed user-item pair, TransRec exploits the translation principle to model the sequential behaviors of users~\cite{recsys2017translation}. In particular, the representation of user $u$ is treated as the translation vector between the representations of the items $i$ and the item $j$ to visit next, namely, $\Mat{q}_{j}+\Mat{p}_{u}\approx\Mat{q}_{i}$.

Distinct from CML that uses simple metric learning that assumes each user's embedding is equally close to every item embedding she likes, LRML introduces the relation vectors $\Mat{r}$ to capture the relationships between user and item pairs~\cite{LRML} .
More formally, the score function is defined as:
\begin{align}
    s_{ui}=\norm{\Mat{p}_{u}+\Mat{e}-\Mat{q}_{i}}^{2}_{F},
\end{align}
where the relation vector $\Mat{e}\in\Space{R}^{d}$ is constructed using a neural attention mechanism over a memory matrix $\Mat{M}$. $\Mat{M}\in\Space{R}^{m\times d}$ is the trainable memory module, hence $\Mat{E}$ is the attentive sum of $m$ memory slots. As a result, the relation vectors not only ensure the triangle inequality, but also achieve better representation ability.

\subsubsection{Neural Network based Metrics}
Distinct from the foregoing that employs linear metrics, recent studies adopt a diverse array of neural architectures, spanning from MLP, Convolutional Neural Network (CNN), and AE as the main building block to mine complex and nonlinear patterns of user-item interactions.

Researchers made attempts to replace similarity modeling between users and items with MLPs, as MLPs are general function approximators to model  any complex continuous function.  NCF is proposed to model the interaction function between each user-item pair with MLPs as: $\hat{r}_{ui}=f_{\text{MLP}}(\Mat{p}_{u}||\Mat{q}_{i})$. Besides, NCF also incorporates a generic MF component into the interaction modeling, thereby making use of both linearity of MF and non-linearity of MLP to enhance recommendation quality. 

Researchers also proposed to leverage CNN based architecture for interaction modeling. This kind of models first generate interaction maps via outer product of user and item embeddings, explicitly capturing the pairwise correlations between embedding dimensions~\cite{ONCF,wsdm2018caser}.
These CNN based CF model focuses on higher-order correlations among representation dimensions. However, such improvements on performance come at the cost of increasing model complexity and time cost.

Besides,  a line of research exploits AEs to fulfill the blanks of user-item interaction matrix directly in the decoder part~\cite{AutoRec,Liang:2018:VAC,strub2015collaborative,CDAE,zhang2017autosvd++,Zhuang:2017:RLP,zhuang2017representation}.
As the encoder and decoder can be implemented via neural networks, such stacks of nonlinear transformations give the recommenders more capacity to model the user representation from complex combinations of all historically interacted items.

\wx{\textbf{Summary}: Many recent studies have shown the superiority of GNNs in the representation learning of users and items. We ascribe the success to (1) the essential data structure, where the user-item interactions can be naturally represented as a bipartite graph between user and item nodes; and (2) GNNs can explicitly encode the crucial collaborative filtering signal of user-item interactions through information propagation process.
As for interaction modeling, compared with the complex functions and metrics, simple inner product is much more efficient especially in the online and large-scale recommendation.}

\section{Content-enriched Recommendation}
\label{sec:content}



\wx{In collaborative filtering, item representations encode the collaborative signal --- behavioral patterns of users --- solely, but ignore the semantic relatedness. To enhance the representation learning, many researchers go beyond the user-item interactions and exploit auxiliary data.}
The auxiliary data could be classified into two categories: content based information and context-aware data. Specifically, the first category of content information is associated with users and items, including general user and item features, textual content~(a.k.a, item tags, item textual descriptions and users' reviews for items),  multimedia descriptions~(a.k.a, images, videos, and audio information), user social networks, and knowledge graphs. In contrast, contextual information shows the environment when users make item decisions, which usually denotes descriptions that beyond users and items~\cite{TKDE2005Survey}. Contextual information includes time, location, and specific data that are collected from sensors~(such as speed,  and 
weather), and so on. Due to page limits, we discuss the most typical contextual data: temporal data. In the following of the two sections, we would give a detailed summary of the content-enriched recommendation and context-aware recommendation.
For the content-enriched recommendation, we classify the related work into five categories based on the available content information: the general features of users and items, the textual content information, the multimedia information, social networks and knowledge graphs.

\begin{table*}
    \caption{Classification of modeling feature-enhanced CF}
    \centering
        \begin{tabular}{|c|c|c|}\hline
           Category & Modeling Summarization & Models \\ \hline
           \tabincell{c}{Second order}  & \tabincell{c}{Model second order correlations \\with embedding based similarity}  & \tabincell{c}{FM~\cite{FM}, FFM~\cite{ffm}}\\ \hline
           \tabincell{c}{\multirowcell{2}{MLP based \\higher order }} & \tabincell{c}{Design better initialization techniques \\to facilitate MLP modeling} & \tabincell{c}{NFM~\cite{NFM}, FNN~\cite{FNN}, PNN~\cite{PNN}, \\DeepCrossing~\cite{DeepCross},~\cite{Covington:2016}} \\  \cline{2-3}
           & Combine deep and shallow features &Wide\&Deep~\cite{cheng2016wide}, DeepFM~\cite{DeepFM}\\ \hline
           Up to $K^{th}$ order modeling & Deep cross network structure for defined order depth & DCN~\cite{Wang:2017:DCN}, xDeepFM~\cite{xDeepFM} \\ \hline
           Tree structure  & Tree enhanced embedding for attentive cross feature aggregation &TEM~\cite{TEM} \\ \hline
        \end{tabular}
    \label{tab:cf_feature}
\end{table*}

\subsection{Modeling General Feature Interactions}

Factorization Machine~(FM) provides an intuitive idea of feature interaction modeling~\cite{FM}. As features are usually sparse, FM first embeds each feature $i$ into a latent embedding $\mathbf{v}_i$, and models second-order interaction of any two 
feature instances with $x_i$ and $x_j$ as: $\mathbf{v}^T_i\times \mathbf{v}_jx_ix_j$. Naturally, FM models the second-order interactions, and reduces the parameter size of computing similarity of any two features with embedding based models. FM has been extended to field-aware FM by expanding each feature with several latent embeddings based on the field  aware property~\cite{ffm}, or higher-order FMs by directly expanding 2-order interactions with all feature interactions~\cite{HFM}. Despite the ability to model higher-order interactions, these models suffer from noisy feature interactions in the modeling process.

Researchers have explored the possibility of adopting neural models to automatically discover complex higher-order feature interactions for CTR prediction and recommendation. As shown in Table~\ref{tab:cf_feature}, besides  FM based approaches, current related work on this topic can be classified into three categories: implicit MLP structures and explicit up to K-th order modeling, and tree enhanced models.

\textbf{MLP based High Order Modeling.} As the feature interactions are hidden, researchers proposed to first embed each feature with an embedding layer, and then exploit MLPs to discover high order correlations. This category can be seen as modeling feature interactions in an implicit way as MLPs are black-box approaches, and we do not know what kind of feature interactions from the output of the MLP structure models.  Since MLPs suffer from training difficulties, some researchers proposed pretraining techniques~\cite{FNN}. Others injected specific structures in MLPs for better capturing feature interactions. DeepCrossing designed residual structures to add back the original input after every two layers of MLPs~\cite{DeepCross}. The NFM architecture has a proposed bi-interaction operation before MLP layers~\cite{NFM}. PNN modeled both the bit-wise interactions of feature embedding interactions and vector-wise feature interactions~\cite{PNN}. Besides the complex high order interactions, another effective approach is to combine the MLP based high order modeling with the classical linear models~\cite{cheng2016wide,DeepFM}.

\textbf{Cross Network for K-th Order Modeling.} The cross network differs from the MLP based approaches with a carefully designed cross network operation, such that a $K^{th}$ layer cross network models the up to $K^{th}$ order feature interactions. The $k^{th}$ hidden layer output $\mathbf{x}_k$ is calculated by the cross operation as: $\mathbf{x}_k=\mathbf{x}_0\mathbf{x}_{k-1}\mathbf{w}_k+\mathbf{b}_k+\mathbf{x}_{k-1}$~\cite{Wang:2017:DCN}. Instead of operating cross operations at a bitwise level, xDeepFM applies cross interactions at the vector-wise level explicitly~\cite{xDeepFM}. These kinds of models are able to learn bounded-degree feature interactions.

\textbf{Tree Enhanced Modeling.} As trees can naturally show cross feature interactions, researchers incorporated trees as a proxy for recommendation with cross feature explanation.
Specifically, TEM~\cite{TEM} first utilizes decision trees to extract high order interaction of features in the form of cross features, and then input embeddings of cross features into an attentive model to perform prediction. As a result, the depth of the decision tree determines the maximum degree of feature interactions. Furthermore, by combining embedding and tree based models seamlessly, TEM is able to unify their strengths --- strong representation ability and explainability.


\subsection{Modeling Textual Content}
Neural network technique has revolutionized Natural Language Processing~(NLP)~\cite{otter2020survey,lappin2021deep}.  These neural NLP models enable multi-level automatic representation learning of textual content, and can be combined in the recommendation framework for better user and item semantic embedding learning. Given the above neural based NLP models, we discuss some typical textual enhanced recommendation models based on the above techniques. Textual content input for recommendation could be classified into two categories: the first category is the content descriptions associated with either items or users, such as the abstract of an article, or the content descriptions of a user. The second category links a user-item pair, such as users annotating tags to items, or writing reviews for products. For the second category, most models summarize the associated content with each user, and each item~\cite{wsdm2017joint,sun2020dual}. Under such a situation, the second category of content information degenerates to the first category. In the following, we do not distinguish the input content data types, and summarize the related work for modeling contextual content into the following categories: autoencoder based models, word embeddings, attention models, and text explanations for recommendation.

%
%
%
%
%
%
%

\textbf{Autoencoder  based Models.} 
By treating item content as raw features, such as bag-of-words and item tag representations, these models use autoencoders and their variants to learn the bottleneck hidden content representations of items~\cite{CDL,wang2016collaborative,dong2017hybrid,li2017CVA,ying2016collaborative,wei2017collaborative,ijcai2017tag,Zhuang:2017:RLP,zhuang2017representation,strub2016hybrid}. 
For example, Collaborative Deep Learning~(CDL)~\cite{CDL} is proposed to simultaneously learn each item $i$'s embedding $\mathbf{q}_i$ as a combination of two parts: a hidden representation from the item content $\mathbf{x}_i$ with a stacked denoising autoencoder and an auxiliary embedding $\mathbf{\theta}$ that is not encoded in the item content as:

\begin{equation} \label{eq:enc_content}
\mathbf{q}_i= f_{e}(\mathbf{x}_{i})+ \mathbf{\theta}_i,\quad  \theta_i\sim \mathrm{N}(0, \sigma^2)
\end{equation}

\noindent where $f_{e}(\mathbf{x})$ transforms raw content input into a bottleneck hidden vector with an autoencoder. $\mathbf{\theta}_i$ is a free item latent vector that is not captured in the item content, which is similar to many classical latent factor based CF models.  In the model optimization process, the objective function is to simultaneously optimize the rating based loss from users' historical behavior and the content-reconstruction loss from the autoencoder:

\begin{equation}
\mathcal{L}= \mathcal{L}_{R}(\mathbf{R}, \mathbf{\hat{R}})+\lambda \mathcal{L}_{X}(\mathbf{X}, f_{d}(f_{e}(\mathbf{X})) ),
\end{equation}

\noindent where $\lambda$ is a parameter that measures the relative weight between the two loss terms. In the above optimization function, $\mathbf{R}$ is the user-item rating matrix and $\mathbf{\hat{R}}$ denotes the predicted rating. Similarly, $\mathbf{X}$ is the item content input and $f_{d}(f_{e}(\mathbf{X}))$ is an reconstructed content from an autoencoder that encodes the item content into a bottleneck representation $f_{e}$, and then reconstructs it with a decoder $f_{d}$.

Following this basic autoencoder based recommendation model, some studies proposed improvements to consider the uniqueness of the content information. 
For example, 
instead of learning a deterministic vector representation of the item content, a Collaborative Variational AutoEncoder~(CVAE) is
proposed to simultaneously recover the rating matrix and the side content information with a variational autoencoder~\cite{li2017CVA}. Researchers also proposed to leverage the item neighbor information from item content to better represent the bottleneck representation of the item~\cite{wsdm2019gated}. 
For some recommendation scenarios, items are also associated with category information. A denoising autoencoder with weak supervision is proposed to learn the distributed representation vector of each item~\cite{okura2017embedding}.  
Besides, as both users and items could be associated with content information, dual autoencoder based recommendation models have been proposed~\cite{Zhuang:2017:RLP,zhuang2017representation,dong2017hybrid,AAAI2019deeply}.

\begin{figure*}
    \centering
	\includegraphics[width=0.95\linewidth]{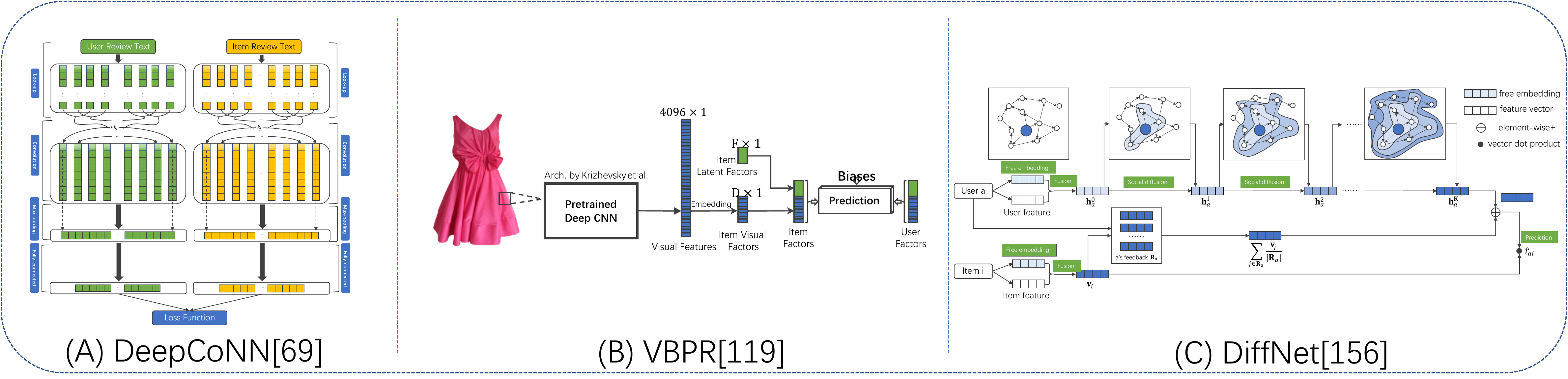}
	\caption{The classical methods for content-enriched models}
	\vspace{-10pt}
	\label{fig:content_method}
\end{figure*}

\textbf{Leveraging Word Embeddings for Recommendation.} Autoencoders provide general neural solutions for unsupervised feature learning, which do not take the uniqueness of text input into consideration. Recently, researchers proposed to leverage word embedding techniques for better content recommendation~\cite{kim2016convolutional,wsdm2017joint,recsys2017transnets,ebesu2017neural,lee2016quote,wu2017joint,fan2019metapath,liu2020kred}. With the success of TextCNN~\cite{kim2014convolutional}, a Convolutional Matrix Factorization (ConvMF) is proposed to integrate CNN into probabilistic matrix factorization~\cite{kim2016convolutional}. Let $\mathbf{x}_i$ denote the text input of item $i$.
The item latent embedding matrix $\mathbf{Q}$ is then represented as a Gaussian distribution that centers around its embedding representation as:

\begin{small}
\vspace{-0.2cm}
\begin{equation} \label{eq:ConMF}
p(\mathbf{Q}|\mathbf{W}, \mathbf{X}, \sigma^2)=\prod_{i=1}^{|V|}\mathcal{N}(\mathbf{q}_i|TextCNN(\mathbf{W},\mathbf{x}_i),\sigma^2) ,
\end{equation}
\vspace{-0.4cm}
\end{small}

\noindent where {\small$\mathbf{W}$} is the parameters in TextCNN. Besides CNN based models, researchers also employed various state-of-the-art content embedding techniques, such as RNNs for item content representation~\cite{bansal2016ask}.

Reviews widely appear in recommendation applications and are natural forms for users to express feelings about items.  Given user's rating records and associated reviews, most review based recommendation algorithms aggregate historical review text of users~(items) as user content input $D(u)$~(item content input $D(i)$). DeepCoNN~\cite{wsdm2017joint} is a deep model for review based recommendation. As shown in Fig.~\ref{fig:content_method}, DeepCoNN consists of two parallel TextCNNs for content modeling: one focuses on learning user behaviors by exploiting review content $D(u)$ written by user $u$, and the other one learns item embedding from  reviews $D(u)$ written for item $i$. After that, a factorization machine is proposed to learn the interaction between user and item latent vectors.
Specifically, DeepCoNN can be formulated as:

\begin{small}
\vspace{-0.2cm}
\begin{equation}
\hat{r}_{ui}=FM(TextCNN(D_u), TextCNN(D_i)).
\end{equation}
\vspace{-0.4cm}
\end{small}


Many studies have empirically found that the most predictive power of review text comes from the particular review of the target user to the target item. As the associated reviews of a user-item pair are not available in the test stage, TransNet is proposed to tackle the situation when the target review information is not available~\cite{recsys2017transnets}.  TransNet has a source network of DeppCoNN that does not include the joint review $rev_{ui}$, and a target network that models the joint review of the current user-item pair $(u,i)$. Therefore, the target network could approximate the predicted review $\hat{rev}_{ui}$ for the test user-item pair even when users do not give reviews to items.

\textbf{Attention Models.}  Attention mechanism has also been widely used in content enriched recommender systems. Given textual descriptions of an item, attention based models have been proposed to assign attentive weights to different pieces of content, such that informative elements are automatically selected for item content representation~\cite{ijcai2016hashtag, seo2017interpretable,colinghashtag,qin2019duerquiz, wang2020fine,qi2020privacy,lee2020news,A3NCF}. For example, given a tweet, the attention based CNN learns the trigger words in the tweet for better hashtag recommendation~\cite{ijcai2016hashtag}. With the historical rated items of a user, an attention model is proposed to selectively aggregate content representations of each historical item for user content preference embedding modeling~\cite{aaai2019dan,aaai2019dynamic,kdd2019npa}. Given user~(item) collaborative embeddings, and  content based embeddings, attention networks have also been designed to capture the correlation and alignment between these two kinds of data sources~\cite{A3NCF,lu2018coevolutionary}. 
Researchers have also proposed a co-evolutionary topical attention regularized matrix factorization model, with the user attentive features learned from an attention network that combines the user reviews, and the item attentive features learned from an attention network that combines the item reviews~\cite{lu2018coevolutionary}. For review based recommendation, researchers argued that most content based user and item representation models neglected the interaction behavior between user-item pairs, and a dual attention model named DAML is proposed to learn the mutual enhanced user and item representations~\cite{kdd2019daml}.  As item content sometimes is presented in multi-view forms~(e.g., title, body, keywords and so on), multi-view attention networks are applied to learn unified item representations by aggregating multiple representations from different views~\cite{ijcai2018interpretable,wu2019neural,gao2020set}. With both the textual descriptions and the image visual information, co-attention is utilized to learn the correlation between the two modalities for better item representation learning~\cite{ma2018MRM,zhang2017hashtag}.


\textbf{Text Explanations for Recommendation.} Instead of improving recommendation accuracy with content input, there is a growing interest of providing text explanations for recommendation. Current solutions for explainable recommendations with text input can be classified into two categories: \textit{extraction based models} and \textit{generation based models}.

\zk{
\textit{Extraction based models} focus on selecting important text pieces for recommendation explanation. Attention techniques are widely used for extraction based explainable recommendation, with the learned attentive weights empirically showing the importance of different elements for model output~\cite{WWW2018neural,ijcai2018interpretable}. After that, the text pieces with larger attentive weights are extracted as recommendation explanations. 
Despite extracting text pieces from reviews, there exist other methods to extract useful text information for explanation, such as review-level explanations~\cite{WWW2018neural,wang2018reinforcement}.
}



\zk{
With the huge success of language generation techniques~\cite{bahdanau2015neural}, \textit{generation based models} draw more and more attention~\cite{li2017neural,li2019persona,chen2019coatt,li2020generate,chen2020towards,sun2020dual,li2020generate}.
Given both users' rating records and reviews, the key idea of these models is to design an encoder-decoder structure, with the encoder part encodes related embeddings of users and items, and the decoder generates reviews that are similar to the ground truth of the corresponding user-item review text. NRT is a state-of-the-art model that simultaneously predicts ratings and generates reviews~\cite{li2017neural}. By taking the one-hot user representation and item representation, the encoder part outputs the user latent embedding and item latent embedding, the review is generated with an RNN based decoder structure, and the rating is predicted with an MLP structure. 
Since we have both the ground truth rating records and the corresponding records of users, the two tasks of rating prediction and review generation can be trained in a multi-task framework. 
Meanwhile, additional information and more advanced encoder-decoder structures are also applied to explanation generation. 
For example, user and item attributes~\cite{li2019persona,chen2019personalized} are multimodal item data~\cite{www2019multimodal}, which are considered in the encoder. 
Then, an advanced attention selector~\cite{chen2019coatt} is designed in the decoder.

}

\subsection{Modeling Multimedia Content}


With the popularity of multimedia based platforms, visual content based multimedia contents, e.g., images and videos, are the most eye-catching for users. In the following, we introduce related work on modeling multimedia content in recommender systems. For ease of explanation, we summarize the related work on multimedia based recommendation with different kinds of input data in Table~\ref{tab:multimedia_rec}.

%

\begin{table*}
    \caption{Multimedia based recommendation models with different kinds of multimedia input}
    \centering
    \begin{tabular}{|c|c|c|}\hline
        Category & Model Summarization & Models\\   \hline
        \multirow{6}{*}{Image} & CNN content based features& ACF\cite{ACF},VBPR\cite{he2016VBPR},OutfitNet\cite{lin2020outfitnet}  \\ \cline{2-3}
          & \tabincell{c}{Aesthetic based features pretrained from \\a deep aesthetic network} & BDN\cite{yu2018aesthetic} \\\cline{2-3}
          & \tabincell{c}{CNN content based features and the style features \\from feature maps of CNNS }& DMF\cite{wen2016visual} \\\cline{2-3}
          & Fine grained image attributes& SAERS\cite{IJCAI2019semantic}, SNMO\cite{AAAI2017examples}, AIC\cite{yang2019interpretable} \\\cline{2-3}
          & \tabincell{c}{Co-attention networks for learning enhanced \\user and image representation} & UVCAN\cite{www2019coattentive} \\\cline{2-3}
          & GNNs to model visual relationships & PinSage\cite{kdd2018gcn}, HFGN\cite{li2020hierarchical}, TransGec\cite{wu2020transgrec}\\\hline
        Image+ Behavior Time& CNN based temporal content evolution & BDN~\cite{yu2018aesthetic}, \cite{he2016ups} \\ \hline
       \multirow{4}{*}{Image+Text}&  Deep fusion networks to learn unified item representation & \tabincell{c}{DMF\cite{wen2016visual}, GraphCAR\cite{Xu2018GraphCAR}, \\CKE\cite{CKE}, Transnfcm\cite{AAAI2019transnfcm}} \\\cline{2-3}
        & Co-attention networks for learning unified item representation & CoA-CAMN\cite{ma2018MRM}, Co-Attention\cite{zhang2017hashtag} \\\cline{2-3}
        & Multi-task learning model with detailed image attributes&\cite{cardoso2018fashion}\\\cline{2-3}
        & Text generation models by encoding user, item text and image content & VECF\cite{chen2019personalized}, KFRCI\cite{Chen2017PKF}, MRG\cite{www2019multimodal}, \cite{yu2020towards}\\ \hline
        Audio & Learning deep audio features & HLDBN\cite{wang2014improving}, \cite{nips2013music}, \cite{nazari2020recommending}\\ \hline
        \multirow{2}{*}{Video} &  \tabincell{c}{Attention networks to learn video representations from \\multiple image representations} & ACF\cite{ACF}, JIFR\cite{IJCAI2019key}, AGCN\cite{wu2020joint} \\\cline{2-3}
        & GNNs to learn video representation & AGCN\cite{wu2020joint}\\  \hline
        Video+Audio & Deep fusion networks to learn unified item representation & CDML\cite{lee2018video}, \cite{lee2017large} \\ \hline
    \end{tabular}
    \label{tab:multimedia_rec}
\end{table*}

\subsubsection{Modeling Image Information}

The current solutions for image  recommendation can be categorized into two categories: content based models and hybrid recommendation models. Content based models exploit visual signals for constructing item visual representations, and the user preference is represented in the visual space~\cite{mcauley2015image,MM2016contagnet,zhang2017hashtag,wen2016visual,AAAI2017examples,Xu2018GraphCAR,lei2016comparative,liu2017deepstyle,IJCAI2019semantic}.
In contrast, the hybrid recommendation models alleviate the data sparsity issue in CF with item visual modeling~\cite{he2016VBPR,he2016ups,ACF,niu2018neural,wang2017imagePOI,cardoso2018fashion,yu2018aesthetic}.

\textbf{Image Content based Models.} Image content based models are suitable for recommendation scenarios that rely heavily on visual influence~(e.g., fashion recommendation) or new items with little user feedback. As visual images are often associated with text descriptions~(e.g., tags, titles), researchers designed some unpersonalized recommender systems that suggest tags to images~\cite{MM2016contagnet,zhang2017hashtag}. These models apply CNNs to extract image visual information, and content embedding models to get textual embedding. Then, in order to model the correlation between visual and textual information, these models either project text and images into a same space~\cite{wen2016visual}, concatenate representations from different modalities~\cite{AAAI2019transnfcm} or design co-attention mechanism to better describe items~\cite{zhang2017hashtag,ma2018MRM}.

For personalized image recommendation, a typical solution is to project both users and items in the same visual space, with the item visual space derived from CNNs, and the user's visual preference either modeled by the items they like~\cite{mcauley2015image} or a deep neural network that takes the user related profiles as input~\cite{lei2016comparative,Xu2018GraphCAR}. Researchers have also argued that CNNs focus on the global item visual representation without fine-grained modeling. Therefore, some sophisticated image semantic understanding models have been proposed to enhance image recommendation performance ~\cite{IJCAI2019semantic,wen2016visual,AAAI2017examples,yang2019interpretable}. For instance, in order to suggest makeups for people, makeup related facial traits are first classified into structured coding. The facial attributes are then fed into a deep learning based recommendation system for personalized makeup synthesis~\cite{AAAI2017examples}. In some visual based recommendation domains, such as the fashion domain, each product is associated with multiple semantic attributes~\cite{yang2019interpretable,IJCAI2019semantic}. To exploit users' semantic preferences for detailed fashion attributes, a  semantic attributed explainable recommender system is proposed by projecting both users and items in a fine-grained interpretable semantic attribute space~\cite{IJCAI2019semantic}.

\textbf{Hybrid Recommendation Models.} Hybrid models utilize both the collaborative signals and the visual content for recommendation, which could alleviate the data sparsity issue in CF and improve recommendation performance. Some researchers proposed to first extract item visual information as features, and  the item visual features are fed into factorization machines for recommendation. Instead of the inferior performance induced by the two step learning process, recent studies proposed end-to-end learning frameworks for hybrid visual recommendation~\cite{he2016VBPR,he2016ups,IJCAI2019key,ACF,niu2018neural,cardoso2018fashion}. Visual Bayesian Personalized Ranking~(VBPR) is one of the first few attempts that leverage the visual content for unified hybrid recommendation~\cite{he2016VBPR}. In VBPR, each user~(item) is projected into two latent spaces: a visual space that is projected from the CNN based visual features, and a collaborative latent space to capture users' latent preferences. Then, given a user-item pair $(u,i)$ with the associated image $x_i$ , the predicted preference $\hat{r}_{ui}$ is learned by combining users' preferences from two spaces:

\begin{equation}\label{eq:vbpr}
\hat{r}_{ui}=\mathbf{p}^T_u\mathbf{q}_i+ \mathbf{w}^T_u f(CNN(x_i)),
\end{equation}

\noindent where $f(CNN(x_i))$ denotes the item content representation by transforming items from the original visual space $CNN(x_i)$. In this equation, the first term models the collaborative effect with free user latent vector $\mathbf{p}_u$ and item latent vector $\mathbf{q}_i$. The second term models the visual content preference with the item visual embeddings as $f(CNN(x_i)$, and the user visual embedding $\mathbf{w}_u$ in the visual space.

Given the basic idea of VBPR, researchers have further introduced the temporal evolution of visual trends in the visual space~\cite{he2016ups}, or the associated location representation of the image~\cite{niu2018neural}. Instead of representing users' preferences into two spaces, the visual content of the item has been leveraged as a regularization term in matrix factorization based models, ensuring that the learned item latent vector of each item is similar to the visual image representation learned from CNNs~\cite{he2016ups}. Besides learning the CNN content representations for item visual representation, many models have been proposed to consider additional information from the imagery for item visual representation, such as the pretrained aesthetics learned from a deep aesthetic network~\cite{yu2018aesthetic}. As users show time-synchronized comments on video frames, researchers proposed a multi-modal framework to simultaneously predict users' preferences to key frames and
generate personalized comments~\cite{Chen2017PKF}. Compared to review generation models~\cite{li2017neural}, the visual embedding is injected into both the user preference prediction part, as well as each hidden state of the LSTM architecture for better text generation.

Recently, GNNs have shown powerful performance in modeling graph data with heuristic graph convolution~\cite{kipf2016semi,tkde2020surveygraph}. PinSage is one of the first few attempts to apply GNNS for web-scale recommender systems~\cite{kdd2018gcn}. Given an item-item correlation graph, PinSage takes node attributes as input, and iteratively generates node embeddings to learn the graph structure with iterative graph convolutions. 
Researchers also proposed to formulate a heterogeneous graph of users, outfits and items, and performed hierarchical GNNs for personalized outfit recommendation~\cite{li2020hierarchical}.

\subsubsection{Video Recommendation}
Researchers proposed content-based video recommender systems with rich visual and audio information~\cite{lee2017large,lee2018video}. Specifically, these proposed models first extracted video features and audio features, and then adopted a neural network to fuse these two kinds of features with early fusion or late fusion techniques. As these content based video recommendation models do not rely on user-video interaction behavior, they can be applied to new video recommendation without any historical behavior data ~\cite{lee2017large,lee2018video}.  In contrast to the content-based recommendation models, with user-video interaction records, researchers proposed an Attentive Collaborative Filtering~(ACF) model for multimedia recommendation~\cite{ACF}. ACF leverages the attention mechanism with visual inputs to learn the attentive weights to summarize users' preferences for historical items and the components of the item. 

The key idea of ACF is to leverage users' multimedia behavior and explicitly project users into two spaces: a collaborative space and a visual space, such that users' key frame preference could be approximated in visual space. The authors designed a model to discern both the collaborative and visual dimensions of users, and model how users make decisive item preferences from these two aspects~\cite{IJCAI2019key}.

\subsection{Modeling Social Network}

With the emergence of social networks, users like to perform item preferences on these social platforms and share their interests with social connections. Social recommendation has emerged in these platforms, with the goal to model the social influence and social correlation among users to boost recommendation performance. The underlying reason for social recommendation is the existence of social influence among social neighbors, leading to the correlation of users' interests in a social network \cite{tsmc2018cnsr,sun2018attentive,wu2018neural,leHASC,leDiffNET,chen2019efficient,liu2020modelling}. We summarize  social recommendation models into following two categories: the social correlation enhancement and regularization models, and GNN based models.

\textbf{Social Correlation Enhancement and Regularization.} By treating users' social behavior as the social domain and item preference behavior as the item domain, the social correlation enhancement and regularization models tried to
fuse users' two kinds of behaviors from two domains in a unified representation. For each user, her latent embedding $\mathbf{p}_u$ is composed of two parts: a free embedding $\mathbf{e}_u$ from the item domain, and a social embedding $\mathbf{h}_u$ that is similar with social connections in the social domain~\cite{tsmc2018cnsr,leHASC,aaai2019hers,kdd2019socialcontent,chen2019efficient}.  In other words, we have:

\begin{flalign}
&\mathbf{h}_u=g(u, \mathbf{S}) \label{eq:social_embed} \\
&\mathbf{p}_u=f(\mathbf{e}_u,\mathbf{h}_u)\label{eq:social_fuse},
\end{flalign}

\noindent  where $g$ models the social embedding part with the social network structure as input, and $f$ fuses the two kinds of embeddings, such as concatenation, addition or neural networks. Different models vary in the detailed implementation of the social domain representation $\mathbf{h}_u$. For example, it can be directly learned from the social network embedding models~\cite{tsmc2018cnsr}, aggregated from the social neighbors' embedding ~\cite{aaai2019hers,leHASC}, or  transferred from the social domain to item domain with attention based transfer learning models~\cite{chen2019efficient}. Besides, the social network is also utilized as a regularization term in the model optimization process, with the assumption that connected users are more similar in the learned embedding space~\cite{tsmc2018cnsr}.

In the real-world, users' interests are dynamic over time due to users' personal interests change and the varying social influence strengths. Researchers extended the social correlation based model with RNN to model the evolution of users' preferences under dynamic social influences~\cite{wu2018neural,sun2018attentive}.  Specifically, for each user $u$, her latent preferences $\mathbf{h}^t_a$ at time $t$ could be modeled as the transition from her previous latent preference $\mathbf{h}^{t-1}_u$, as well as the social influence from social neighbors at $t-1$ as:

\begin{small}
\begin{equation}
\mathbf{h}^t_u=f_{RNN}(R^t_u, \mathbf{h}^{t-1}_{u}, \sum_{a\in S_u}t_{au}\mathbf{h}^{t-1}_{a})
\end{equation}
\end{small}

\noindent where $R^t_u$ is the temporal behaviors of user $u$ at this time, $\sum_{a\in S_u}t_{au}h^{t-1}_a$ denotes the influences from her social neighbors. In particular, the social influence strength $t_{au}$ could be simply set as equally for each social neighbor, or with attention modeling for influence strength inference.


\textbf{GNN Based Approaches.}  Most of the above social recommendation models utilized the local first-order social neighbors for social recommendation. In the real world, the social diffusion process presents a dynamic recursive effect to influence a user's decision. In other words, each user is influenced recursively by the global social network graph structure. To this end, researchers argued that it is better to leverage the GNN based models to better model the global social diffusion process for recommendation. DiffNet is designed to simulate how users are influenced by the recursive social diffusion process for social recommendation with the social GNN modeling. Specifically, DiffNet recursively diffuses the social influence from step $0$ to the stable diffusion depth $K$. Let $\mathbf{h}^k_u$ denote the user embedding at the $k^{th}$  diffusion process, which is modeled as:

\begin{small}
\vspace{-0.2cm}
\begin{flalign}
&\mathbf{h}^0_u=f_{NN}(\mathbf{x}_u, \mathbf{e}_u) \label{eq:diffnet_u0}\\
&\mathbf{h}^{(k-1)}_{Su} =Pool(\mathbf{h}^{(k-1)}_a| a\in S_u)  \label{eq:diffnet_su}\\
&\mathbf{h}^k_u=s(W^k[\mathbf{h}^{k-1}_{Su},\mathbf{h}^{(k-1)}_u]) \label{eq:diffnet_uk}
\end{flalign}
\end{small}

\noindent where Eq.\eqref{eq:diffnet_u0} fuses the user feature $\mathbf{x}_u$ and user free latent vector  $\mathbf{e}_u$ with a neural network $f_{NN}$ for initial influence diffusion. At each diffusion step $k$, Eq.\eqref{eq:diffnet_su} models the influence diffusion from $u$'s social neighbors, and Eq.\eqref{eq:diffnet_uk} depicts the user embedding at the recursive step $k$ by fusing her previous embedding $\mathbf{h}^{k-1}_u$ and influences from her social neighbors as $\mathbf{h}^{k-1}_{Su}$. As $k$ diffuses from step $1$ to depth $K$, the recursive social diffusion process is captured.

Instead of performing GNNs on the user-user social graph, researchers have also considered jointly modeling the social diffusion process in the social network and the interest diffusion process in the user-item graph with heterogeneous GNN based models~\cite{fan2019gnnsocial,wsdm2019session,wu2019dualgraphatt,leTKDE2020diffnet++,satuluri2020simclusters,liu2020modelling,jin2020partial}. For instance, DiffNet++ is proposed to jointly model the interest diffusion from user-item bipartite graph and the influence diffusion from the user-user social graph for user modeling in social recommendation, and have achieved state-of-the-art performance
~\cite{leTKDE2020diffnet++}.

\subsection{Modeling Knowledge Graph}
Researchers have also considered leveraging Knowledge Graphs~(KG) for recommendation, which provide rich side information for items (\ie item attributes and external knowledge). 
Typically, KG organizes such subject-property-object facts in the form of directed graph $\Set{G}=\{(h,r,t|h,t\in\Set{E},r\in\Set{R})\}$, where each triplet presents that there is a relationship $r$ from head entity $h$ to tail entity $t$.
Exploring such interlinks, as well as user-item interactions, being a promising solution to enrich item profile and enhance the relationships between users and items.
Furthermore, such graph structure endows recommender systems the ability of reasoning and explainability~\cite{DBLP:journals/corr/abs-1906-05237,KPRN,KTUP,KGAT,zhao2020leveraging,wang2020learning,zhu2020knowledge}.
Recent efforts for KG enhanced recommendation can be roughly categorized into three categories: path-based models~\cite{DBLP:conf/wsdm/YuRSGSKNH14,KGRnn18,KPRN,MetaFMG,kdd2018metapath}, regularization-based models~\cite{CKE,Huang:2018:ISR,KTUP,DBLP:conf/www/WangZZLXG19}, and GNN-based approaches~\cite{KGAT,KGCN,tai2020mvin,chen2020jointly,wu2020joint,sun2020multi,zhou2020improving,lee2020news,jin2020multi,li2020hierarchical,hu2020graph,KGIN2020}.

\vspace{5px}
\textbf{Path Based Methods.}
Many efforts introduce meta-paths~\cite{DBLP:conf/wsdm/YuRSGSKNH14,gao2018recommendation,MetaFMG,kdd2018metapath,gong2020attentional,wang2020disenhan,han2020genetic,wang2020reinforced} and paths~\cite{KPRN,KGRnn18,wang2020kerl,symeonidis2020recommending,lei2020interactive} that present high-order connectivity between users and items, and then feed them into predictive models to directly infer user preferences.
In particular, a path from user $u$ to item $i$ can be defined as a sequence of entities and relations: $p=[e_{1}\xrightarrow{r_{1}}e_{2}\xrightarrow{r_{2}}\cdots\xrightarrow{r_{L-1}}e_{L}]$, where $e_{1}=u$ and $e_{L}=i$, and $(e_{l},r_{l},e_{l+1})$ is the $l$-th triplet in $p$, and $L-1$ denotes the number of triplets in the path. As such, the set of paths connecting $u$ and $i$ can be defined as $\Set{P}(u,i)=\{p\}$.

FMG~\cite{MetaFMG}, MCRec~\cite{kdd2018metapath}, and KPRN~\cite{KPRN} convert the path set into an embedding vector to represent the user-item connectivity.
Such paradigm can be summarized as follows:
\begin{align}
    \Mat{c}=f_{\text{Pooling}}(\{f_{\text{Embed}}(p)|p\in\Set{P}(u,i)\}),
\end{align}
where $f_{\text{Embed}}(\cdot)$ embeds path $p$ as a trainable vector. $f_{\text{Pooling}}(\cdot)$ is the pooling operation to synthesize all path information into the connectivity representation, such as the attention networks adopted in MCRec and KPRN. RippleNet~\cite{wang2018ripple} constructs ripple set (\ie high-order neighboring items derived from $\Set{P}$) for each user to enrich her representations.

While explicitly modeling high-order connectivity, it is highly challenging in real-world recommendation scenarios because most of these methods require extensive domain knowledge to define meta-paths or labor-intensive feature engineering to obtain qualified paths~\cite{DBLP:journals/corr/abs-1906-05237,KGAT}. Moreover, the scale of paths can easily reach millions or even larger when a large number of KG entities are involved, making it prohibitive to efficiently transfer knowledge.

\textbf{Regularization Based Methods.} This research line devises a joint learning framework, where direct user-item interactions are used to optimize the recommender loss, and KG triples are utilized as additional loss terms to regularize the recommender model learning. In particular, the anchors between two modeling components are the embeddings of the overlapped items. CKE~\cite{CKE} makes use of Knowledge Graph Embedding (KGE) techniques, especially TransR~\cite{lin2015learning}, to generate additional representations of items, and then integrates them with item embeddings of the recommender MF, which is defined as:
\begin{align}
    \Mat{q}_{i}=f_{\text{Embed}}(i) + f_{\text{KGE}}(i|\Set{G}),
\end{align}
where $f_{\text{Embed}}(\cdot)$ is the embedding function which takes the item ID as the input, while $f_{\text{KGE}}$ is the output of KGE method which considers the KG structure. Similarly, DKN~\cite{DKN} generates item embeddings from NCF and TransE. These approaches focus on enriching item representations by the joint learning framework. 

\textbf{GNN Based Methods.} The regularization-based methods only take direct connectivity between entities into consideration, while encoding the high-order connectivity in a rather implicit manner. Due to the lack of explicit modeling, neither the long-range connectivities are guaranteed to be captured, nor the results of high-order modeling are interpretable~\cite{KGAT}. More recent studies, such as KGAT~\cite{KGAT}, CKAN~\cite{wang2020ckan}, MKM-SR~\cite{meng2020incorporating}, and KGCN~\cite{KGCN}, get inspired by the advances of GNNs and explore the message-passing mechanism over graphs to exploit high-order connectivity in an end-to-end fashion.

KGAT~\cite{KGAT} encodes user-item interactions and KG as a unified relational graph by representing each user behavior as a triplet, ($u$, Interact, $i$).
Based on the item-entity alignment set, the user-item bipartite graph can be seamlessly integrated with KG as a so-called collaborative knowledge graph $\Set{G}=\{(h,r,t)|h,t\in\Set{E}',r\in\Set{R}\}$, where $\Set{E}'=\Set{E}\cup\Set{U}$ and $\Set{R}'=\Set{R}\cup\{\text{Interact}\}$.
Over such graph, KGAT recursively propagates the embeddings from a node's neighbors (which can be users, items, or other entities) to refine the node's embedding, and employs an attention mechanism to discriminate the importance of the neighbors as:
\begin{align}
    \Mat{p}_{u}=f_{\text{GNN}}(u,\Set{G}),
\end{align}
where $f_{\text{GNN}}(\cdot)$ is the GNN component.

\zk{
\textbf{Summary:} 
Auxiliary data, such as text, multimedia, and social network, is capable of enhancing the user and item representation learning and boosting the recommendation performance. 
The keys are the selection of auxiliary data and the integration methods. 
For example, text information can help models to generate corresponding recommendation explanation. 
Social network information is very useful to provide social influence and social correlation among users for better recommendation. 
Meanwhile, attention mechanism is a general method to select the most relevant information from auxiliary data to enhance the representation learning. 
GNN-based methods are good at obtaining structure information and high-order correlation for the utilization of auxiliary data. 
As a conclusion, based on the recommendation target~(recommendation accuracy, explanation, cold-start problem, etc), selecting proper auxiliary data and integration method can help recommendation models to achieve a good performance.
}
\section{Temporal/Sequential Models}
\label{sec:contextRS}

\begin{figure*}
    \centering
	\includegraphics[width=0.90\linewidth]{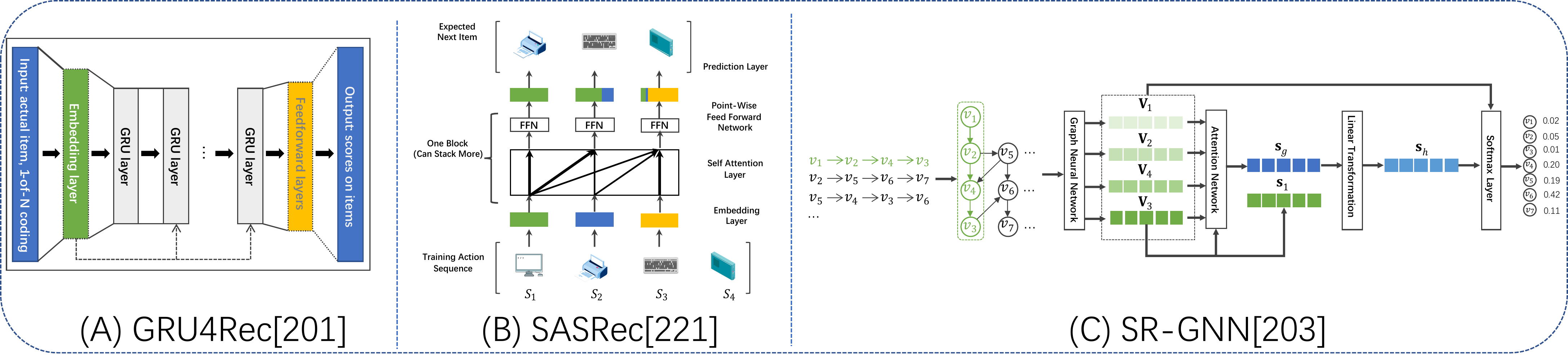}
	\caption{The classical methods for temporal/sequential Models}
	\vspace{-8pt}
	\label{fig:sequence_model}
\end{figure*}

Users' preferences are not static but evolve over time. Instead of modeling users' static preferences with the aforementioned models, temporal/sequential based recommendation focuses on modeling users' dynamic preferences or sequential patterns over time. Given a userset {\small$\mathcal{U}\!=\![u_1,u_2,..., u_M]$}
and an itemset {\small$\mathcal{V}\!=\![i_1,i_2,...,i_N]$}, current temporal/sequential recommendation could be generally classified into three categories:

\begin{itemize}
  \item \emph{Temporal based recommendation}: For a user {\small$u\!\in\!\mathcal{U}$} and an item {\small$i\!\in\!\mathcal{V}$}, the associated user-item interaction behavior is denoted as a quadri-tuple as $[u, i, r_{ui}, t_{ui}]$. In this representation, $r_{ui}$ denotes the detailed rating and $t_{ui}$ is the timestamp of this behavior.  Temporal recommendation focuses on modeling the temporal dynamics of users' behavior over time.
  \item \emph{Session based recommendation}: In a certain session {\small$\mathcal{s}=[i_1,i_2,...,i_{|S|}]$~($s\subseteq \mathcal{V}$)}, a user interacts with a collection of items~(e.g., consumption with a shopping basket, browsing the internet in a limited time period). In many session based applications, users do not log in and user IDs are not available~\cite{hidasi2016session,hidasi2016parallel,aaai2019session}. Therefore, the popular direction of session based recommendation is to mine the sequential item-item interaction patterns from the session data for better recommendation.
  \item \emph{Temporal and session based recommendation}: This approach combines the definition of temporal recommendation and session recommendation, in which each transaction is described as $[u, s, t]$, with $s\subseteq V$ is a collection of items that are consumed at a particular time $t$. Under this scenario, both the temporal evolution and the sequential patterns of items need to be captured.
\end{itemize}
\begin{table*}
    \caption{A comparison of methods that models the  temporal and sequential effects in RS}
    \centering
    \begin{tabular}{|c|c|c|}\hline
        Model Type & Model Summarization & Models     \\   \hline
        \multirow{2}{*}{Temporal Models} & \tabincell{c}{Recurrent neural networks  \\to capture temporal evolution }& \tabincell{c}{ARSE\cite{sun2018attentive},RRN\cite{wsdm2017recurrent}  \\
        \cite{wsdm2017survival}, \cite{dlrs2016rnncoe}, \cite{www2018evolution},\cite{wsdm2018latentcross},\cite{recsys2017sequential}} \\\cline{2-3}
         &   Memory network based models & NMRN \cite{kdd2018memory}, MANN \cite{wsdm2018seqmemory}, STAMP \cite{kdd2018stamp} \\ \hline
        \multirow{6}{*}{Sequential Models}& \tabincell{c}{RNN based models that rely on sessions \\ to construct input and output} &\tabincell{c}{p-RNN\cite{hidasi2016parallel},KERL\cite{wang2020kerl},CRNNs\cite{dlrs2017contextual},\\NARM\cite{CIKM2017neural},\cite{dlrs2016improvedrnn},\cite{recsys2017rnnknn},\cite{rescys2017intent}, \cite{hidasi2016session},} \\\cline{2-3}
          & \tabincell{c}{Translation based models for modeling \\the correlations of consecutive items}&\tabincell{c}{TransRec\cite{recsys2017translation},PeterRec\cite{yuan2020parameter},\cite{yin2020learning}}\\\cline{2-3}
          & Convolutional sequence embedding models & 3D CNNs\cite{recys20173DCNN} \\\cline{2-3}
          &Self attention for learning item correlations &SASRec\cite{icdm2018SASRec}, MFGAN\cite{ren2020sequential} \\\cline{2-3}
          & \tabincell{c}{Memory network to learn the session representation}  &DMN\cite{sigir2018mobile} \\\cline{2-3}
          & \tabincell{c}{GNN based models for learning \\item correlations}&\tabincell{c}{SR-GNN\cite{aaai2019session},GC-SAN\cite{IJCAI2019graph},Gag\cite{qiu2020gag},\\GCE-GNN\cite{wang2020global},SGNN-HN\cite{pan2020star},\cite{chen2020handling}}
          \\
        \hline
        \multirow{6}{*}{\tabincell{c}{Temporal and \\Sequential Models}} &  \tabincell{c}{Hierarchical attention networks with \\long and short term interest}&\tabincell{c}{SHAN\cite{ijcai2018hierar},HRM\cite{wang2015learning},HGN\cite{kdd2019hiesequential},\\MARank\cite{aaai2019multi},Fissa\cite{lin2020fissa},SSE-PT\cite{wu2020sse}} \\\cline{2-3}
       & RNN based models&RRN\cite{kdd2018hispre},BINN\cite{wsdm2017recurrent},HIERNN\cite{recsys2017hiernn} \\\cline{2-3}
      & \tabincell{c}{Attention based models for \\user interest modeling} &\tabincell{c}{CTRec\cite{bai2019ctrec},M3\cite{tang2019towards},S3-rec\cite{zhou2020s3},CTA\cite{wu2020deja},\\MTAM\cite{ji2020sequential},TASER\cite{ye2020time},ReChorus\cite{wang2020make}}\\\cline{2-3}
      & \tabincell{c}{Memory Networks for long distance item correlations}&KA-MemNN\cite{zhu2020sequential},CSRM\cite{wang2019collaborativesession},MTAM\cite{ji2020sequential} \\\cline{2-3}
     &  CNN based models&Caser\cite{wsdm2018caser},CTRec\cite{bai2019ctrec},\cite{WSDM2019simple}         \\ \cline{2-3}
     & GNN based models& HyperRec\cite{wang2020next},MA-GNN\cite{ma2020memory},IMfOU\cite{guo2020intention}  \\
    \hline
    \end{tabular}
    \label{tab:time_rs}
\end{table*}
\noindent We summarize the main techniques for modeling temporal and sequential effects in recommender systems in Table~\ref{tab:time_rs} and illustrate some representative work in Fig.~\ref{fig:sequence_model}.

\subsection{Temporal based recommendation} 
Temporal recommendation models focus on capturing the temporal evolution of users' preferences over time. Due to the superior of RNNs in modeling temporal patterns, many temporal based approaches take RNNs into consideration. Recurrent Recommender Networks~(RRN) is one of the representative studies for temporal recommendation by  endowing both users and items with an LSTM autoregressive architecture~\cite{wsdm2017recurrent}. In RRN, the predicted rating $\hat{r}^t_{ui}$ of user $u$ to item $i$ at time $t$ is modeled as:

\begin{small}
\vspace{-0.2cm}
\begin{flalign}
\hat{r}^t_{ui}&=f(\mathbf{p}^t_u, \mathbf{q}^t_i) \quad \mbox{where} \label{eq:rrn_predr} \\
\mathbf{p}^{t}_{u}&=RNN(\mathbf{p}^{(t-1)}_{u}, \mathbf{W}\mathbf{x}^t_{u}),\quad
\mathbf{q}^{t}_{i}=RNN(\mathbf{q}^{(t-1)}_i, \mathbf{W}\mathbf{x}^t_{i}) \label{eq:rrn_uv}
\end{flalign}
\end{small}

\noindent where $\mathbf{p}^t_u$  and $\mathbf{q}^t_i$ are the dynamic embeddings of user $u$ and item $i$ at time $t$, respectively. Specifically, $f$ in Eq.\eqref{eq:rrn_predr} is a temporal rating prediction function.
Eq.\eqref{eq:rrn_uv} models the evolution of users and items' dynamic embeddings with RNN architecture. 
As the user side and item side share similar LSTM structure, we take the user side as an example. $\mathbf{x}^t_{u}\in \mathbb{R}^{|V|} $ is a rating vector for $u$ between $t-1$ and current time $t$, with each element $x^t_{ul}$ denotes the rating of user $u$ to  each item $l$ at that time.  $\mathbf{W}$ is a transformation matrix that needs to be learned. Therefore, RRN learns the evolution of user and item latent vectors over time with two RNNs. 
Based on RRNs, rich context factors were considered, such as the social influence~\cite{sun2018attentive,www2018evolution}, item metadata~\cite{wsdm2018latentcross,tkde2021ekt} and multimedia data fusion~\cite{tkde2020mvrnn}. Take the RNN in the user side as an example, and we can generalize the user latent embedding evolution as:

\begin{equation}
\mathbf{p}^t_u=RNN(\mathbf{p}^{(t-1)}_{u}, \mathbf{W}\mathbf{x}^t_{u}, Contextual Embedding),
\end{equation}

\noindent where additional contextual embeddings are also injected to model  temporal evolution of users' temporal embedding.

Recently, an emerging trend is to model the temporal evolution with Neural Turning Machines~\cite{ntm2014neural} and Memory Networks~\cite{nips2015mn}.  Compared to RNNs, memory networks introduce a memory matrix to store the states in memory slots, and update memories over time with read and write operations.  As the memory storage is limited, the key component in applying memory networks in recommendation is how to update memories over time with users' temporal behavior. Researchers proposed a general memory augmented neural network with user memory networks to store and update users' historical records, and the user memory network is implemented from the item and feature level~\cite{wsdm2018seqmemory}. Researchers further proposed to use attention mechanism in the memory reading and writing process with soft-addressing, in order to better capture users' long-term stable and short-term temporal interests~\cite{kdd2018memory}.

\subsection{Session based recommendation}  
Many real-world recommender systems often
encounter the short session data from anonymous users,  i.e., the user ID information is not available. Session based recommendation is popular under this situation, which models the sequential item transition patterns given many session records.
Hidasi et al.~\cite{hidasi2016session} made one of the first few attempts to design GRU4REC for session based recommendation under the RNN based framework.
Specifically, GRU4REC resembles an RNN structure, which recursively takes the current item in the session as input, updates the hidden states, and outputs the predicted next item based on the hidden state. Given anonymous sessions, the key component of GRU4REC is how to construct mini-batches to suit the data forms of RNNs. Since the goal is to capture how a session evolves over time with item dependencies, the authors designed a session parallel min-batches. The first events of the first several sessions are extracted to form the first mini-batch, with the desired output is the second event of the corresponding session. Under such a formulation, the complex correlations of items in a session are captured for session based recommendation.

GRU4REC has been further investigated with item feature consideration~\cite{hidasi2016parallel}, local intent~\cite{CIKM2017neural}, user information consideration~\cite{quadrana2017personalizing}, data augmentation techniques~\cite{dlrs2016improvedrnn}. By treating item ID, name, and category with an embedding matrix, a sequence of clicks could be represented as frames.  
Therefore, the architecture of 3D CNNs could be transferred to session-based recommendation~\cite{recys20173DCNN}. 
Furthermore, a self attention based sequential model of SASRec is proposed. SASRec models the entire user sequence without any recurrent and convolutional operations, and adaptively considers consumed items for recommendation~\cite{icdm2018SASRec}. 


Researchers also proposed a translation based model to capture the personalized sequential third order interactions between a user $u$, the previous item $j$, and the current item $i$.
Given the item embedding matrix $\mathbf{Q}$, each user's embedding $\mathbf{p}_u$ can be approximated as:
$\mathbf{q}_i+\mathbf{p}_u\approx \mathbf{q}_j$~\cite{recsys2017translation}. Therefore, the translation based models capture the correlation of two constructive items.

While above models built relationships between consecutive items in a session, how to globally model the transitions in a session among distant items remain under explored. Researchers adopted GNNs for session based recommendation~\cite{aaai2019session,IJCAI2019graph,qiu2020gag,wang2020global,chen2020handling,pan2020star,DBLP:conf/aaai/WuT0WXT19,DBLP:conf/ijcai/XuZLSXZFZ19}.  SR-GNN is one of the first few attempts. 
As shown in Fig.~\ref{fig:sequence_model}, the graph is constructed by taking all items as the graph node set, and there is an edge between two nodes if these two nodes appear in consecutive orders in a session. Then, the GNN is adopted to learn item embeddings, such that the higher-order relationships of items from session behavior data can be modeled~\cite{aaai2019session}. Different GNN based models vary in graph construction, and graph aggregation process~\cite{IJCAI2019graph,qiu2020gag,wang2020global}.

\subsection{Temporal and session based recommendation}
Given the session data of each user over time,  models in this category leverage both the temporal evolution modeling of users, as well as the sequential item patterns hidden in the sessions for recommendation.  Currently, the solutions could be classified into two categories: the first category learns both users' long term preference and the short term dynamic preferences, and the second category adopts advanced neural models for learning a unified user representation.

In the first category, each user's long term preference is modeled from her historical behaviors, and the short term dynamics is modeled from the previous session or the current session~\cite{wang2015learning,ijcai2018hierar,aaai2019multi} . For example, researchers proposed hierarchical attention networks for temporal and session based recommendation, with the first attention layer learns the user long term preference based on historical records, and the second one attentively aggregates user representation from the current session as:

\begin{equation}
\mathbf{p}^t_u=Att_2(\mathbf{p}_u, Att_1(\mathbf{q}_l,l\in T^{(t-1)}_{u})),
\end{equation}

\noindent where $Att_1$ denotes the bottom layer attention network that depicts the user's short term preference from recent user behavior $T^{(t-1)}_{u}$, and $Att_2$ is a top layer attention network that balances the short term user preference and long term preference embedding vector $\mathbf{p}_u$.  Instead of using hierarchical attentions, researchers proposed to adopt attention techniques to learn item correlations, and designed recurrent states at top layers for sequential recommendation~\cite{wu2020deja}.

Hierarchical RNNs are also proposed for personalized session-based recommendation over time, with a session level GRU unit to model the user activity within sessions, and a user level GRU models the evolution of the user preference over time~\cite{recsys2017hiernn,meng2020incorporating}. Besides, researchers exploited hierarchical attention networks to learn better short term user preference with feature-level attention and item level attention~\cite{kdd2019hiesequential}.  For the long term user interest modeling, researchers proposed to leverage nearby sessions~\cite{kdd2018hispre}, designed attention modeling or memory addressing techniques to find related sessions~\cite{bai2019ctrec,KDD2019dualsequential,zhu2020sequential,wang2019collaborativesession}.

Another kind of models utilize the 3D convolutional networks for recommendation, which defines the recommendation problem as~\cite{wsdm2018caser,WSDM2019simple}: $(S^u_{t-L}, ...., S^u_{t-2}, S^u_{t-1}) \rightarrow S^u_t$, where $S^u_t\subseteq V$ is the $t$-th time sequential behavior of user $u$ at time $t$, and $L$ denotes the maximum sequence length. Convolutional Sequence Embedding Recommendation~(Caser) is a representative work that incorporates CNNs to learn the sequential patterns. It captures both user's general preferences and sequential patterns, at both the union level and point level with convolution operations, and captures the skip behavior~\cite{wsdm2018caser,WSDM2019simple}. 

Besides, researchers proposed to leverage the advances of GNN based models for recommendation~\cite{wang2020next,guo2020intention,ma2020memory}. The graph structure is constructed from all sessions to form a global item correlation graph or graphs at each time period. For example, researchers constructed time-aware hypergraphs to model item correlations over time. After that, the self attention modules are used to model users' dynamic interests based on the learned dynamic item embeddings over time~\cite{wang2020next}.

\zk{
\textbf{Summary:}
Temporal/sequential based models focus on the dynamic preferences of users over times. 
Therefore, most existing work concentrates on the sequential information of users and items, and leverages sequential models (e.g., RNN, Memory Network) to capture the trends of user preference evolution. 
The main challenges lie in the recognition of long-term and short-term temporal interests, as well as the identification of global and local interests in the absence of user ID information. 
Since GNNs are skilled at processing user-item interactions at different granularities, we can observe that it receives more and more attention in temporal/sequential based models.
}

%
%
%

\section{Discussion and Future Directions} \label{sec:conclusion}
\zk{
The foregoing various neural network based recommendation models have demonstrated the superior recommendation quality. 
However, we realize that current solutions for recommendation are far from satisfactory, and there are still many opportunities in this area. 
Therefore, we outline some possible directions that deserve more research efforts from the basis, modeling, and evaluation perspectives. 
Last but not least, we present a discussion about the reproducibility of recommendation models. 
}


\textit{Basis: Recommendation Benchmarking.} While the field of neural recommender systems has seen a great surge of interests in recent years, it has also been difficult for researchers to keep track of what represents the state-of-the-art models. It is urgent to identify the architectures and key mechanisms that generalize to most recommender models.  However, this is a non-trivial task as recommendation scenarios are diverse, e.g., static recommendation models or dynamic recommendation models, content enriched or knowledge enhanced models. Different recommendation models rely on different data sets with varying inputs. Besides, the same model would have varying performance on different recommendation scenarios due to the assumption in the modeling process. In fact, the Netflix competition for CF based recommendation has passed more than 10 years, how to design a large benchmarking recommendation dataset that keeps track of the state-of-the-art recommendation problems and update the leading performance for comparisons is a challenging yet urgent future direction.

\textit{Models: Graph Reasoning $\&$ \wx{Self-supervised Learning}.} Graphs are ubiquitous structures in representing various recommendation scenarios.
For instance, CF could be seen as a user-item bipartite graph, content based recommendation is represented as an attributed user-item bipartite graph or a heterogeneous information network~\cite{tkde2016surveyhin,tkde2020surveygraph}, and knowledge enhanced recommendation is defined as a combination of knowledge graph and user-item bipartite graph. With the great success of deep learning on graphs~\cite{tkde2020surveygraph}, it is promising to design graph based models for recommendation. Some recent studies have empirically demonstrated the superiority of graph embedding based recommendation models, how to explore the natural graph reasoning techniques for better recommendation is a promising direction.
\wx{Besides, self-supervised learning \cite{DBLP:conf/iclr/GidarisSK18,DBLP:journals/corr/abs-1807-03748} is becoming emerged and showing promises in recommendation tasks \cite{SGL,yang2021enhanced,DBLP:journals/corr/abs-2107-05315,DBLP:conf/aaai/0013YYWC021,chen2021set2setrank}. Its core is to distill extra supervision signals from the limited available user interaction data via some auxiliary tasks and facilitate the downstream recommendation tasks. As such supervisions are complementary to the user-item interactions, they enhance the representation learning of users and items.
Incorporating self-supervised learning into recommendation could offer promising solutions to the long-standing issues of data sparsity and long-tail distribution.
}



\textit{Evaluation: Multi-Objective Goals for Social Good Recommendation.
} Recommender systems have penetrated every aspect in our daily life, and have greatly shaped the decision process of providers and users. Most previous recommender systems concentrated on the single goal of recommendation accuracy based user experience. These systems limit the ability to incorporate user satisfaction from multiple goals, e.g., recommendation diversity and explanations to persuade users~\cite{zhang2018explainable}.  Besides, the user-centric approach neglects system objectives from multistakeholders and the society. The data-driven approaches with accuracy as goals may lead to biases in the algorithmic process decision process~\cite{mehrabi2019fair,chen2020bias,www2021wu,fcs2021FairCF}. For recommender systems, researchers have realized that long tailed items have fewer chances to be recommended,  and benefiting users may obscure concerns that might come from other stakeholders in this system. How to provide multi-objective goals for social good recommendation, such as explainability, balance of multistakeholders, and fairness for the society is an important research topic that needs to be paid attention to.

\wx{\textit{Discussion: Reproducibility.} While the neural recommendation models have dominated in the recommendation field and claimed substantial improvements over previous models, recent efforts  raise questions about their reproducibility and published claims~\cite{DBLP:conf/sigir/JiSZL20,recsys2019worryanalysis,DBLP:journals/tois/DacremaBCJ21,DBLP:conf/recsys/RendleKZA20,rendle2019difficulty}.
This can be attributed to two aspects. First, neural recommendation models are based on neural networks, which are hard to tune in practice. Thus, we should carefully choose the initialization, tune hyperparameters, avoid model collapse, and so on. Besides, due to the various application scenarios of recommendation, different models vary in the selection of datasets and setting of experiments. Specifically, it is well known that recommender models are sensitive to the dataset size, the dataset sparsity, the data preprocessing and splitting techniques, the strategy of negative sampling, the choice of loss function and optimization manner, and the evaluation metrics of performance. \zk{
Thus, it is very challenging to conduct a fair performance comparison.
In order to advance the recommendation community, some researchers make efforts on the data level, such as industry-relevant recommendation benchmark~\cite{wu2020developing}, MIcrosoft News Dataset (MIND)~\cite{wu2020mind}, and Yelp dataset\footnote{\url{https://www.yelp.com/dataset}}. 
Others concentrate on the unified evaluation framework~\cite{paudel2019mix,said2014rival}. 
}For example, researchers argue that previously default choice of evaluating recommender models with sampled metrics (e.g., rather than using the full set, only sampling a small set of negative items during testing) would be inconsistent to the true trend \cite{DBLP:conf/kdd/KricheneR20}.
Towards fair and reproducible comparisons, it is of crucial importance to make the experimental settings transparent (e.g., release the codes, datasets, and experimental settings, and set up a leaderboard if possible).
Furthermore, beyond network architecture engineering and hunting for the ``best'' performance, research studies on theoretical considerations and reproducibility analysis should be encouraged.
}

\zk{
\section{Conclusion}
\label{s:conclusion}
In this survey, we provide a systematic review on neural recommender models from the perspective of recommendation modeling with accuracy goal.
Based on the data usage, we organize existing work into three categories: \textit{collaborative filtering model}, \textit{content enriched model}, and \textit{temporal/sequential model}. 
In each part, we summarize a bunch of influential research work and conclude corresponding main contributions as well as our opinions. 
Moreover, we also elaborate possible promising directions from the basics, modeling, and evaluation perspectives, and reproducibility problem in recommender systems. 
Still, a large number of novel methods and techniques are proposed each year. 
We hope this survey is able to help reader to quickly understand the development and key aspects of recommendation modeling, and inspires some future studies.
}


\ifCLASSOPTIONcaptionsoff
  \newpage
\fi

\bibliographystyle{IEEEtran}

\bibliography{reference}

\vspace{-8mm}
\begin{IEEEbiography}[{\includegraphics[width=1in,height=1.25in,clip,keepaspectratio]{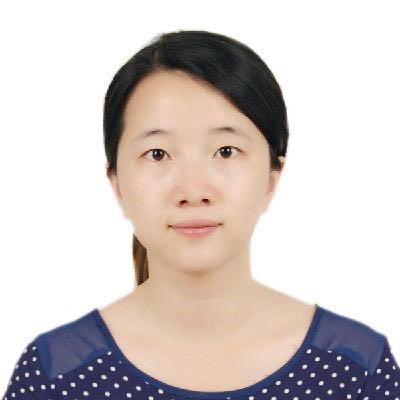}}]{Le Wu}
	is currently an associate professor and Ph.D. supervisor at the Hefei University of Technology (HFUT), China. She received the Ph.D. degree from the University of Science and Technology of China (USTC). Her general area of research interests is data mining, recommender systems and social network analysis. She has published more than 50 papers in referred journals and conferences. Dr. Le Wu is the recipient of the Best of SDM 2015 Award, the Distinguished Dissertation Award from China Association for Artificial Intelligence (CAAI) 2017, and the Youth Talent Promotion Project from China Association for Science and Technology.
\end{IEEEbiography}
\vspace{-8mm}
\begin{IEEEbiography}[{\includegraphics[width=1in,height=1.25in,clip,keepaspectratio]{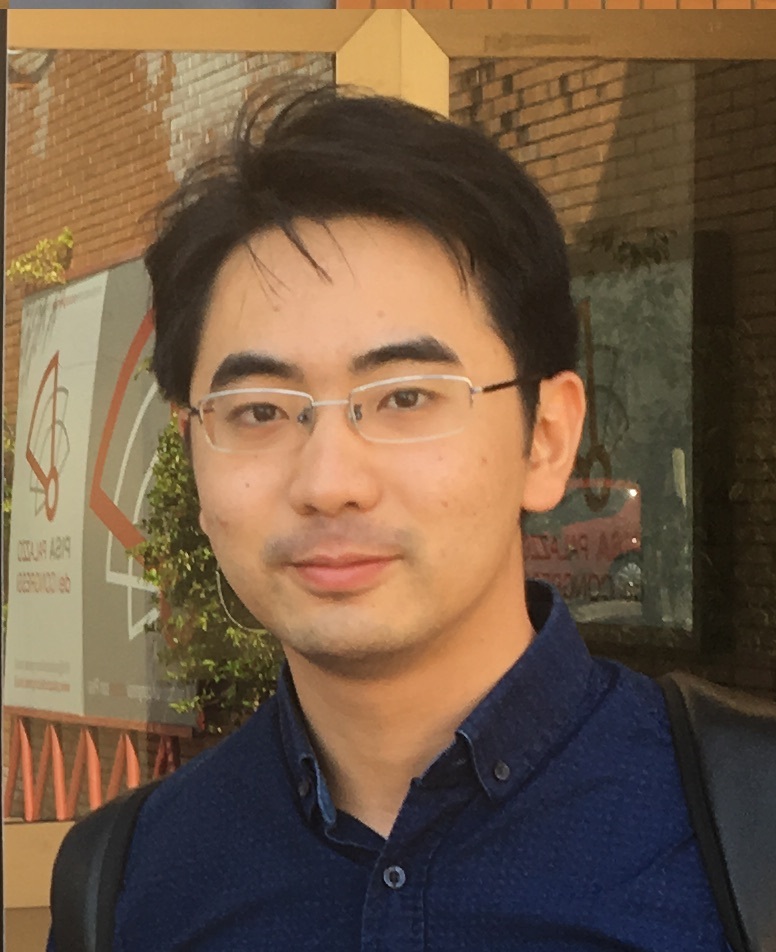}}]{Xiangnan He}
	is a professor at the University of Science and Technology of China (USTC). He received his Ph.D. in Computer Science from the National University of Singapore (NUS). His research interests span information retrieval, data mining, and multi-media analytics. He has over 80 publications that appeared in several top conferences such as SIGIR, WWW, and MM, and journals including TKDE, TOIS, and TMM. His work has received the Best Paper Award Honorable Mention in WWW 2018 and ACM SIGIR 2016. He is in the editorial board of journals including Frontiers in Big Data, AI Open etc. Moreover, he has served as the PC chair of CCIS 2019 and SPC/PC member for several top conferences including SIGIR, WWW, KDD, MM, WSDM, ICML etc., and the regular reviewer for journals including TKDE, TOIS, TMM, etc.
\end{IEEEbiography}
\vspace{-8mm}
\begin{IEEEbiography}[{\includegraphics[width=1in,height=1.25in,clip,keepaspectratio]{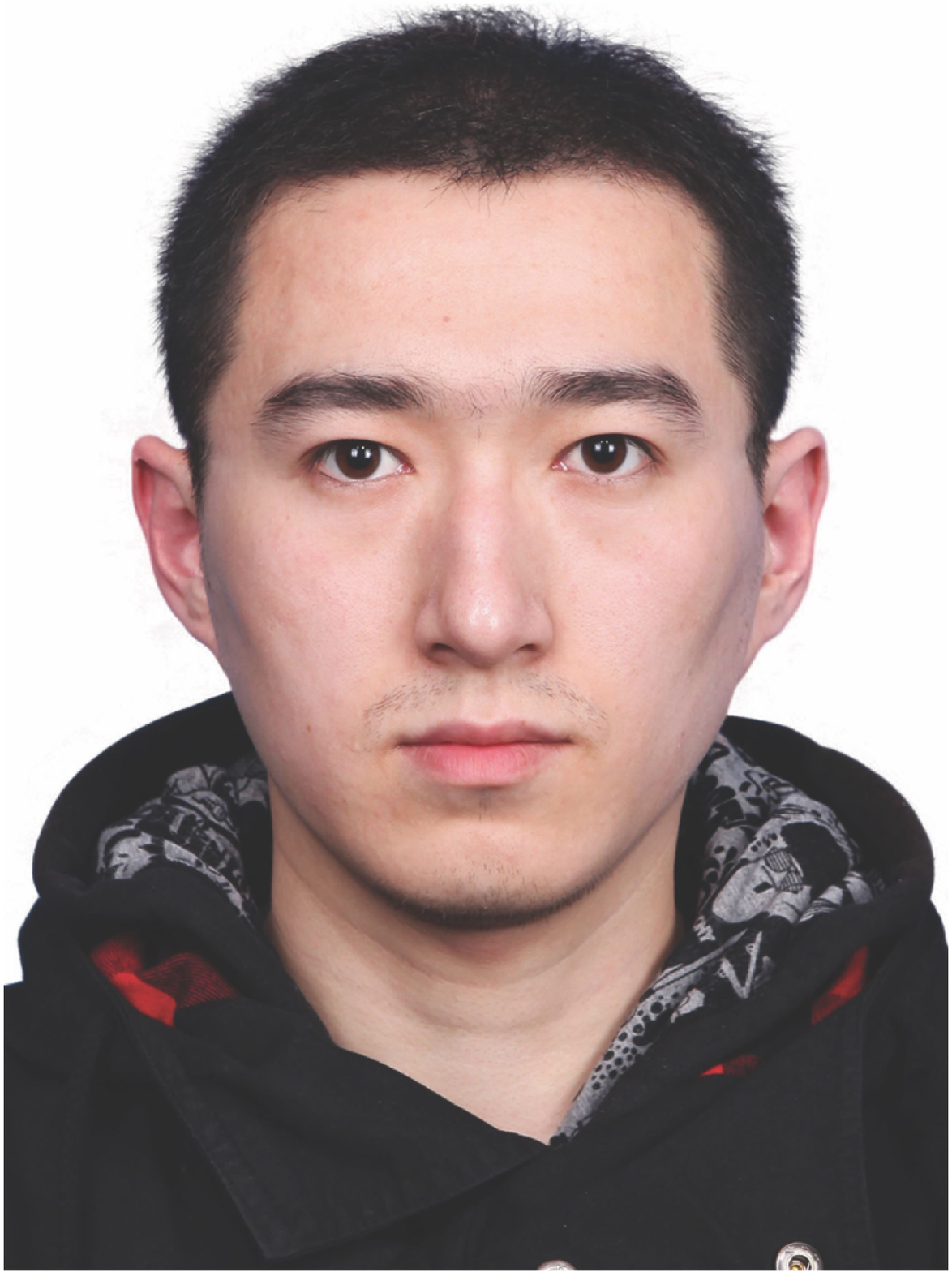}}]{Xiang Wang}
	 is now a research fellow at National University of Singapore. He received his Ph.D. degree from National University of Singapore in 2019. His research interests include recommender systems, graph learning, and explainable deep learning techniques. He has published some academic papers on international conferences such as KDD, WWW, SIGIR, and AAAI. He serves as a program committee member for several top conferences such as KDD, SIGIR, WWW, and IJCAI, and invited reviewer for prestigious journals such as TKDE, TOIS, TNNLS, and TMM.
\end{IEEEbiography}

\begin{IEEEbiography}[{\includegraphics[width=1in,height=1.25in,clip,keepaspectratio]{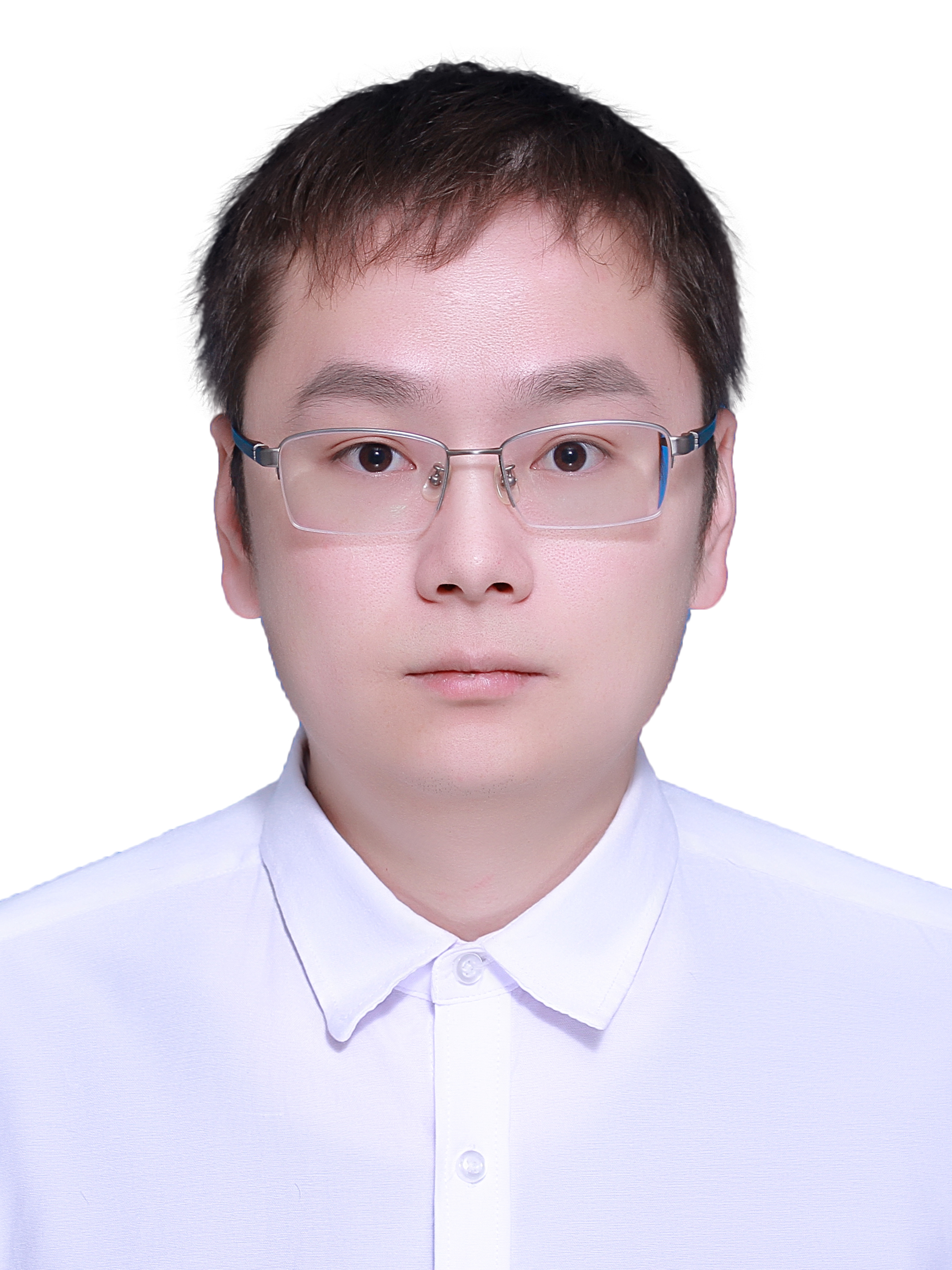}}]{Kun Zhang}
	 received the PhD degree in computer science and technology from University of Science and Technology of China, Hefei, China, in 2019. He is is currently a faculty member with the Hefei University of Technology (HFUT), China. His research interests include Natural Language Understanding, Recommendation System. He has published several papers in refereed journals and conferences, such as the IEEE Transactions on Systems, Man, and Cybernetics: Systems, the ACM Transactions on Knowledge Discovery from Data, AAAI, KDD, ACL, SIGIR, WWW, ICDM. He received the KDD 2018 Best Student Paper Award.
\end{IEEEbiography}
\vspace{-30mm}
\begin{IEEEbiography}[{\includegraphics[width=1in,height=1.25in,clip,keepaspectratio]{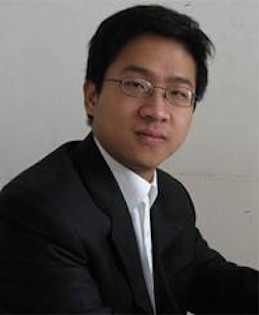}}]{Meng Wang} received the BE and PhD degrees from USTC, in 2003 and 2008, respectively. He is a professor with HFUT. His current research interests include multimedia content analysis, computer vision, and pattern recognition. He has authored more than 200 book chapters, journal, and conference papers in these areas. He is the recipient of the ACM SIGMM Rising Star Award 2014. He is an associate editor of the IEEE Transactions on Knowledge and Data Engineering, the IEEE Transactions on Circuits and Systems for Video Technology, and the IEEE Transactions on Neural Networks and Learning Systems. He is an IEEE Fellow and IAPR Fellow.
\end{IEEEbiography}

\end{document}